\documentclass[aps,prb,twocolumn,superscriptaddress,showpacs]{revtex4}
\usepackage{graphicx,epsfig}
\usepackage{amssymb,times}
\usepackage{amsmath,amsfonts,bm}

\newcommand{\aver}[1]{{\langle #1\rangle}}
\newcommand{\mubar}{\overline{\mu}}
\begin{document}

\title{Thermodynamics of the three-dimensional Hubbard model:\\
Implications for cooling cold atomic gases in optical lattices}

\author{Lorenzo De Leo}
\affiliation{Centre de Physique Th\'eorique, Ecole Polytechnique, CNRS,
91128 Palaiseau, France}
\author{Jean-S\'ebastien Bernier}
\affiliation{Centre de Physique Th\'eorique, Ecole Polytechnique, CNRS,
91128 Palaiseau, France}
\author{Corinna Kollath}
\affiliation{Centre de Physique Th\'eorique, Ecole Polytechnique, CNRS,
91128 Palaiseau, France}
\affiliation{Departement of Theoretical Physics, University of Geneva, 24 
Quai Ernest Ansermet, 1211 Gen\`eve, Switzerland}
\author{Antoine Georges}
\affiliation{Centre de Physique Th\'eorique, Ecole Polytechnique, CNRS,
91128 Palaiseau, France}
\affiliation{Coll\`ege de France, 11 place Marcelin Berthelot, 75005 Paris, France}
\author{Vito W. Scarola}
\affiliation{Department of Physics, Virginia Tech, Blacksburg,
Virginia 24061, USA}

\begin{abstract}
We present a comprehensive study of the thermodynamic properties of the three-dimensional
fermionic Hubbard model, with application to cold fermionic atoms subject to an
optical lattice and a trapping potential.
Our study is focused on the temperature range of current experimental interest.
We employ two theoretical methods - dynamical mean-field theory and high-temperature
series - and perform comparative benchmarks to delimit their respective range of
validity.
Special attention is devoted to understand the implications that thermodynamic properties
of this system have on cooling.
Considering the distribution function of local occupancies in the inhomogeneous
lattice, we show that, under adiabatic evolution, the variation
of any observable (e.g., temperature) can be conveniently disentangled into two distinct contributions.
The first contribution is due to the redistribution of atoms in the trap during the evolution, while
the second one comes from the intrinsic change of the observable.
Finally, we provide a simplified picture of a recently proposed cooling procedure, based on
spatial entropy separation, by applying this method to an idealized model.

\end{abstract}

\date{\today}

\maketitle

\section{Introduction}

One of the main ongoing efforts in cold atom gases is the investigation of strongly
correlated phases.
Our theoretical understanding of strongly correlated phases is, in general, far from complete.
For example, computational studies, of fermions in particular, are severely limited.  
As a result, the possibility of performing analog simulations of model
Hamiltonians, using cold atoms in optical 
lattices\cite{JakschZoller1998, GreinerFoelling2008, Cho2008, BlochZwerger2008, esslinger_annurev_2010},
has raised great hopes.
The remarkable controllability of cold atom systems, allowing, for example, the
application of a specific time-dependent perturbation, has also opened the possibility of studying
strongly correlated systems in regimes inaccessible to solid state materials,
especially away from equilibrium.

Among all model Hamiltonians relevant to the physics of strong correlations,
the Hubbard model has attracted the greatest attention.
On one hand, this model is the simplest to have important competition between 
kinetic and potential energies. It also plays in the field of strong
correlations a somewhat analogous role to the one played by the Ising model in classical
statistical mechanics. On the other hand,
its physics is of direct relevance to high-temperature superconductivity,
a phenomenon still veiled in mystery.
In fact, this new field often dubbed `condensed matter of light and atoms' was pioneered by the
theoretical prediction\cite{JakschZoller1998} and experimental observation of
an incompressible regime, characteristic of a Mott
insulator, and of the transition between this phase and itinerant ones, first
for bosons\cite{GreinerBloch2002} and recently also
for fermions\cite{JoerdensEsslinger2008,SchneiderRosch2008}.
For recent reviews, see, e.g., Refs.~\onlinecite{BlochZwerger2008, esslinger_annurev_2010}.

In order to make progress towards the ultimate goal of performing analog simulations of model
Hamiltonians, a good synergy between experimental efforts and theoretical investigations is
crucial.
For example, theoretical inputs are useful in establishing maps, in parameter space,
of the location of the different phases present in a realistic setup which takes
into account the trap confining potential.
Moreover, experiments are currently confronted with the great difficulty
of cooling fermions in optical lattices to sufficiently low temperatures to reach 
many interesting strongly correlated phases.
This relevant temperature range is significantly lower than the one corresponding
to mere quantum degeneracy.
Theoretical control over these issues is greatly needed and requires a quantitative
understanding of the thermodynamic properties of the Hubbard model, both for the homogeneous system and
in the presence of a trap.

In this article, we perform a comprehensive study of the thermodynamic properties
of the homogeneous three-dimensional fermionic Hubbard model, and of cold fermionic atoms
in a three-dimensional optical lattice subjected to a trapping potential. We focus on
the range of temperature which is of direct interest for current experiments as well as for the
next generation of experiments. Particular emphasis is put on aspects related to
cooling of the system.

The main theoretical technique used in the present study is dynamical mean-field
theory (DMFT). This approach (reviewed, e.g., in Ref.~\onlinecite{GeorgesRozenberg1996})
is a controlled approximation which is able to capture the competition
between the kinetic energy, that tends to
delocalize atoms over the whole lattice, and the repulsive potential energy
that prevents atoms from occupying the same site, hence promoting localization.
We also use another theoretical approach, namely high order
high-temperature series expansions\cite{HendersonAshley1992, tenHaafvanLeeuwen1992, Oitmaa2006, ScarolaTroyer2009}.
One of the goals of the present article is to delimit, in parameter space, the respective range of
validity of each of these approaches, and to provide a
benchmark for their use through quantitative comparisons.

This article is organized as follows. In Sec.~\ref{sec:model_methods}, we define the
model considered in this article, specify notations and conventions and briefly outline
the theoretical methods used.
In Sec.~\ref{sec:hom}, we provide detailed results for the thermodynamics of the
homogeneous Hubbard model in three dimensions.
Finally, the effect of the trapping potential is considered in Sec.~\ref{sec:trap},
with several applications geared towards cooling fermionic cold atom systems.

\section{Model, Methods and Conventions}
\label{sec:model_methods}

Cold fermionic gases with two hyperfine states loaded in an optical lattice
potential can be described in a wide range of parameters by the fermionic
Hubbard model\cite{JakschZoller1998,HofstetterLukin2002,WernerHassan2005}. The
model Hamiltonian is given by
\begin{eqnarray} \label{eq:Hubb_Ham}
    H&=& -J \sum_{\langle j,j'\rangle,\sigma} \left(c_{j,\sigma}^\dagger
    c^{\phantom{\dagger}}_{j',\sigma}+h.c.\right)
    + U \sum _{j} \hat{n}_{j,\uparrow} \hat{n}_{j,\downarrow} \nonumber\\
    &&  -\, \mu \sum_{j,\sigma}\,\hat{n}_{j,\sigma}
    +\,  \sum_{j,\sigma}V(r_j)\,\hat{n}_{j,\sigma}.
\end{eqnarray}
where $c_{j,\sigma}^\dagger$ is the creation operator of a fermion with spin
$\sigma$ at site $j$ and $\hat{n}_{j,\sigma} = c_{j,\sigma}^\dagger
c^{\phantom{\dagger}}_{j,\sigma}$ is the density operator at site $j$. $r_j$ is
measured in units of the lattice spacing. The spin $\sigma = \downarrow, \uparrow$
labels the two hyperfine states of the atoms, and $\langle j,j'\rangle$ labels neighboring 
sites.

The first term in the Hamiltonian describes the kinetic energy of the atoms
with $J$ the hopping amplitude between nearest-neighbor sites.
In this paper, we mainly consider a three dimensional
cubic lattice and use the half-bandwidth $6J$ as our energy units.
The second term represents
the $s$-wave scattering between fermions in different hyperfine states, and its
strength $U$ is proportional to the s-wave scattering length which can be tuned over a wide range using a Feshbach resonance.
We often use a grand-canonical description of the system
in which the filling can be adjusted by the chemical
potential $\mu$. The last term in the Hamiltonian represents the external trapping potential.
Aspects related to this trapping term will be discussed in more detail later.

We investigate the properties of the system in a moderately
high temperature regime, typically $k_B T/6J \gtrsim 1/10$, or $\beta 6J\lesssim 10$
where $\beta\equiv 1/k_BT$ is the inverse temperature. This is the temperature regime
of interest in current ultracold quantum gases experiments\cite{JoerdensTroyer2010}.
Also one should note that we often use $k_B=1$.

At moderate temperature and commensurate filling (one atom per site) the
homogeneous model displays a crossover between a liquid phase at weak
interaction and an incompressible regime at larger interaction. The latter is
characterized by suppressed density fluctuations and is a Mott insulator.
The properties of these two phases will be discussed in detail
in the next sections. A simple picture of these two regimes can already be obtained by
considering the extreme limits of vanishing interaction and vanishing hopping.
In the first case, the system only consists of a mixture of non-interacting free
fermions. The fermions are delocalized and form a Fermi sea. In the case of
vanishing hopping, the so-called atomic limit, the system consists of
disconnected sites, and the problem of a single site can be solved. The atoms localize and a 
Mott insulator forms at half filling. 

The atomic limit can be
seen as the zeroth order term in a high-temperature expansion of the grand potential, 
$\Omega = - \ln{(\mathcal{Z})}/\beta $, in $\beta J$, where $\mathcal{Z}=\text{Tr} \exp(-\beta H)$ 
is the grand partition function.  The series for the grand potential in a uniform system ($V=0$) can be written:
\begin{equation}
-\beta \tilde{\Omega}=\ln z_{0}+\sum_{m=2}^{\infty} (\beta J/z_{0})^{m}X^{(m)}(w,\zeta ),
\label{grand}
\end{equation}
where  $\zeta = \exp(\beta\mu)$ is the fugacity, $ \tilde{\Omega}\equiv\Omega/N$, $w=\exp(-\beta U)$ 
and $ z_{0}=1+2\zeta+\zeta^2w$ is the partition function of a single site in the atomic limit.  The atomic 
limit can be improved by systematically calculating\cite{ScarolaTroyer2009,HendersonAshley1992, 
tenHaafvanLeeuwen1992, Oitmaa2006,Kubo1980,KuboTada1983,KuboTada1984}  higher order
series coefficients, $X^{(m)}$.  Thermodynamic quantities at high temperatures can be accurately 
derived from leading terms in Eq.~\ref{grand}.

For intermediate couplings and temperatures, the solution of the Hubbard model
is highly non trivial.
In this article, we mainly use dynamical mean-field theory (DMFT)\cite{GeorgesRozenberg1996} to explore this regime.
DMFT is an approximate method in which the lattice model is replaced by
a self-consistent impurity model. This model can be solved using highly accurate numerical algorithms
such as the strong-coupling CT-QMC algorithm\cite{WernerMillis2006} which we use in the
present work. Although approximate, DMFT is a controlled approximation: it is exact in both
non-interacting and atomic limits, and bridges the gap between the two limits. Furthermore,
it becomes mathematically exact in the formal limit of infinite lattice coordination.
Physically, DMFT neglects spatial correlations, but treats accurately local quantum
fluctuations.
Besides DMFT, we also use high-order high-temperature series
expansions \cite{ScarolaTroyer2009,HendersonAshley1992, tenHaafvanLeeuwen1992, Oitmaa2006,Kubo1980,KuboTada1983,KuboTada1984}.
In fact, one of the aims of this article is to provide benchmarking of DMFT in the high-temperature domain
from comparison with these series. More details on the regions of validity of the two methods are
given in Sec.~\ref{sec:theory_validity}.

Finally, at low temperature ($\beta6J\gtrsim 16$),
a phase transition into an antiferromagnetically order phase occurs\cite{StaudtMuramatsu2000,WernerHassan2005},
this regime will not be considered in the rest of this paper.

\section{Homogeneous Hubbard model}
\label{sec:hom}

In this section we focus on the properties of the homogeneous Hubbard model.
We discuss in particular the density, the double occupancy and the entropy in this model.
These quantities are very important in the characterization
of ultra cold quantum gases. The effect of the trap will be
considered in Sec.~\ref{sec:trap}.

\subsection{Density, double occupancy and entropy}

\begin{figure}[bt!]
    \includegraphics[width=7cm,clip=true]{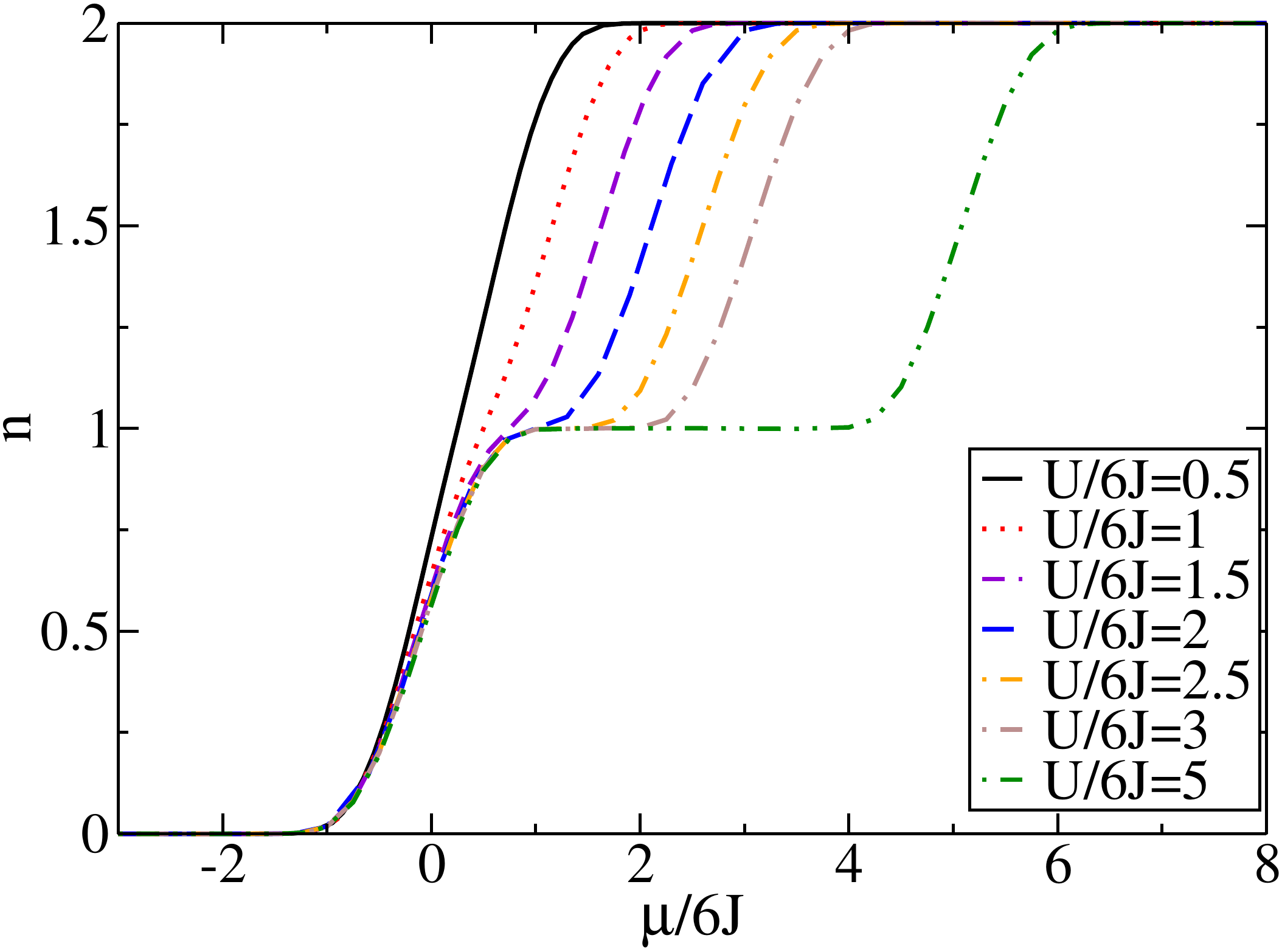}
    \caption{(Color online)  Density $n$ versus chemical potential $\mu$ for different interaction strength $U$
             and at fixed inverse temperature $\beta6J=10$ (DMFT). For large interaction strength a clear Mott plateau of filling $n=1$ develops.
       \label{fg:n_mu_low_T}}
\end{figure}

The physics of the competition between kinetic and potential energy in the
Hubbard model is exemplified by the behavior of the density per site $n_j =
\langle n_{j,\uparrow} + n_{j,\downarrow} \rangle$ as a function of the
chemical potential $\mu$. For simplicity, since we are considering the homogeneous system, we drop the site index in this section. At low enough
temperature, this quantity has a qualitatively different behavior for weak and
strong interaction (Fig.~\ref{fg:n_mu_low_T}). At weak coupling (cf.~$U/6J<1$
in Fig.~\ref{fg:n_mu_low_T}), $n(\mu)$ has a smooth evolution from $n=0$
for $\mu \ll -6J$ to $n=2$ for $\mu \gg 6J+U$. In this case, the compressibility
$\kappa = \partial n / \partial \mu$ is always finite when $0<n<2$. In the opposite
limit of strong coupling (cf.~$U/6J>2$ in Fig.~\ref{fg:n_mu_low_T}), the potential energy disfavors the
presence of two fermions on a single site and $n$ develops a plateau around $\mu = U/2$.
On this plateau, characteristic of the Mott insulating state, the
compressibility $\kappa$ vanishes in the zero temperature limit. As the chemical potential
must be of the order of $U$ to overcome the potential interaction and
to generate doubly occupied sites, the central plateau widens with increasing
interaction.

\begin{figure}[bt!]
    \includegraphics[width=7cm,clip=true]{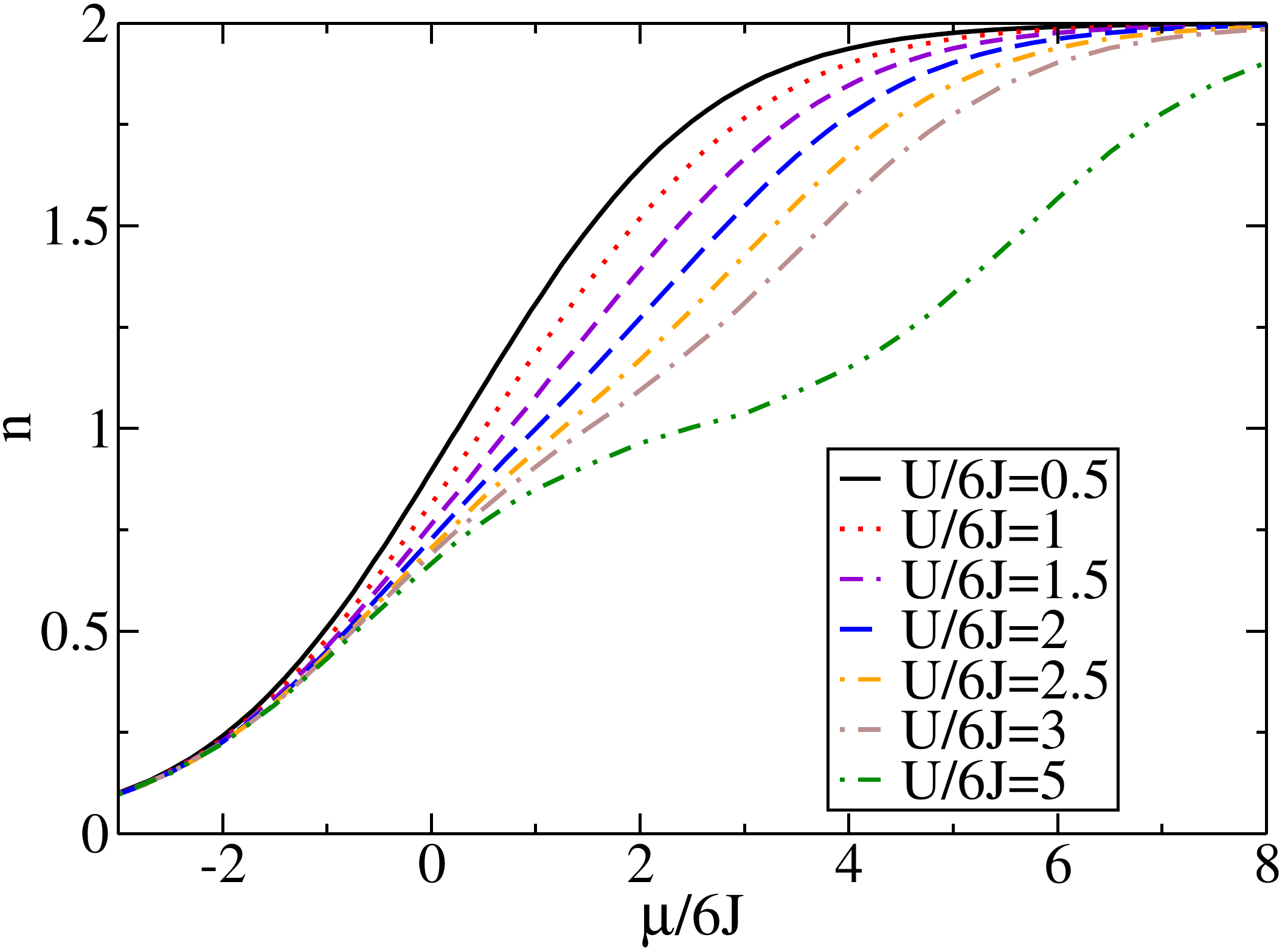}
    \caption{(Color online) Density $n$ versus chemical potential $\mu$ for increasing interaction $U$ and fixed inverse temperature $\beta 6J =1$ (DMFT).
             Only a reminiscent behavior of the Mott plateau is left for large interaction strength.
    }
    \label{fg:n_mu_high_T}
\end{figure}

Increasing the temperature causes a softening of these features. Fig.~\ref{fg:n_mu_high_T} shows the density plotted at $k_BT/6J=1$.
%
%
At weak interaction strength, thermal excitations reduce the slope which characterizes
the variation of the density with chemical potential. For strong interactions, large temperatures
of the order of $U$ (the Mott gap) are needed to generate thermal excitations and to destroy the Mott
plateau. One can see comparing Figs.~\ref{fg:n_mu_low_T} and \ref{fg:n_mu_high_T} how the
plateau found at $k_BT/6J=0.1$ for $U/6J>2.5$ disappears at high temperature only leaving behind
an inflection in the $U/6J=5$ curve.

\begin{figure}[bt!]
    \includegraphics[width=7cm,clip=true]{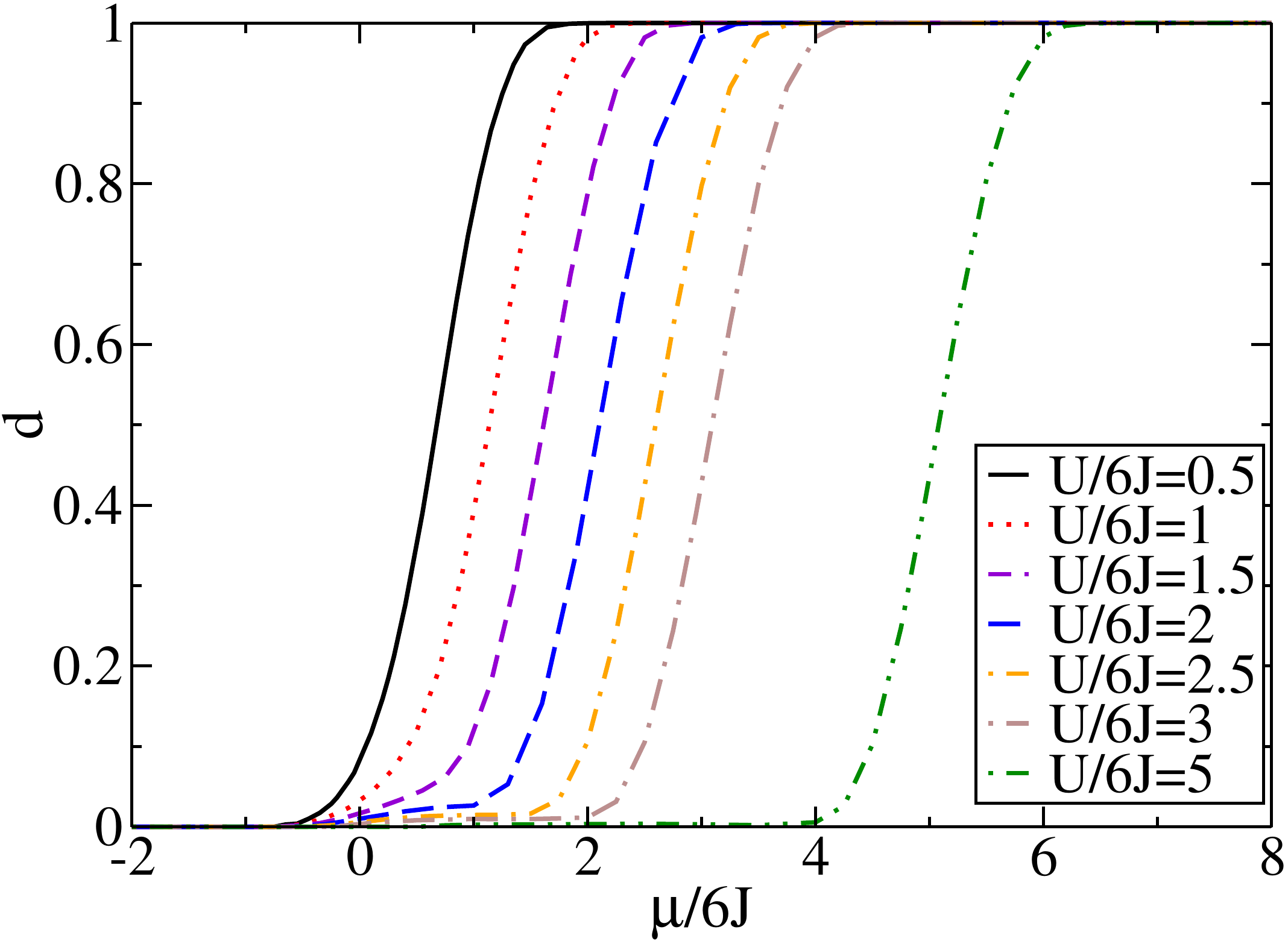}
    \caption{(Color online) Double occupancy $d$ versus chemical potential $\mu$ at fixed inverse temperature $\beta 6J=10$ and different values of the interaction strength $U$ (DMFT).
    }
    \label{fg:D_mu_beta10}
\end{figure}

\begin{figure}[bt!]
    \includegraphics[width=7cm,clip=true]{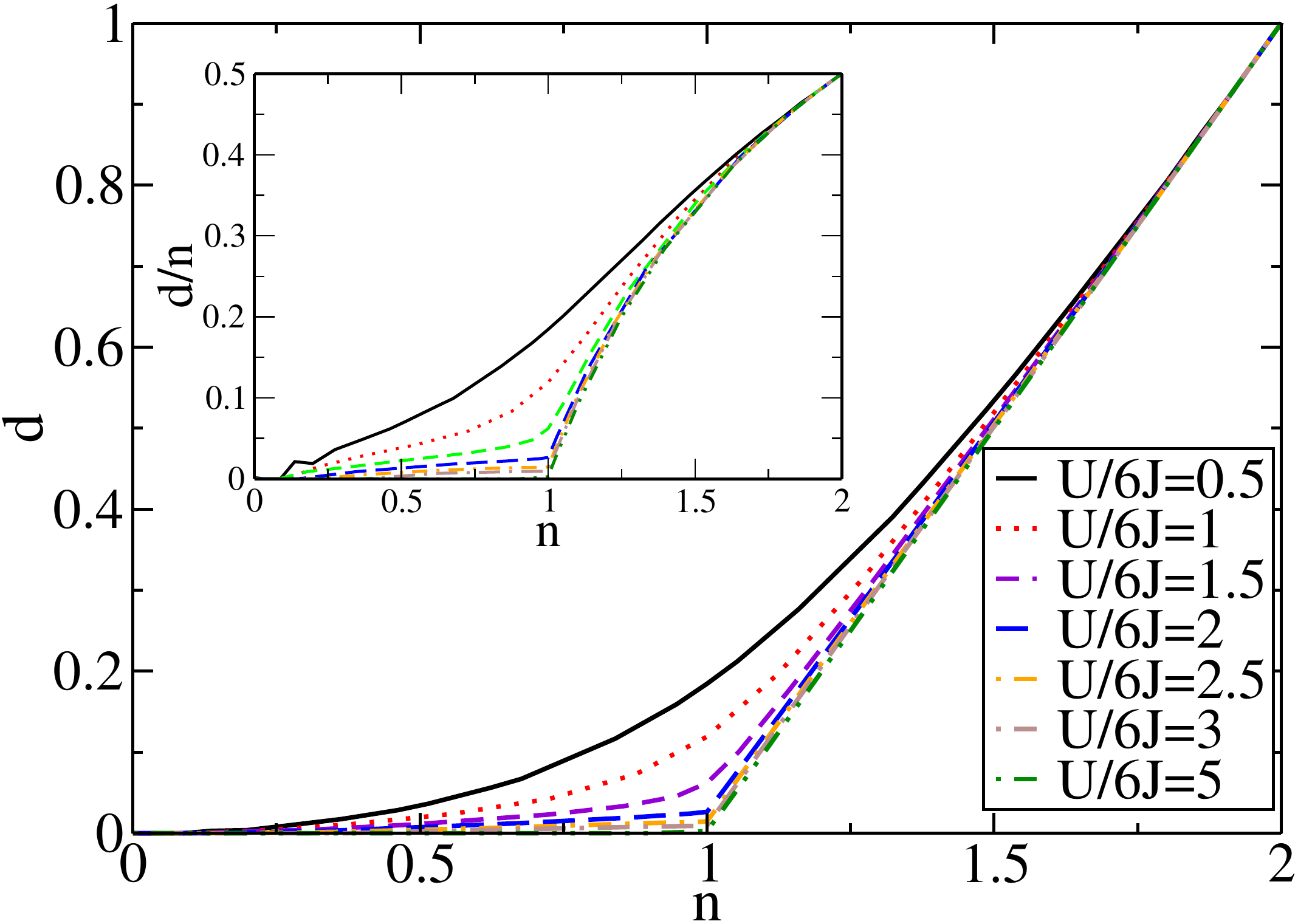}
    \caption{(Color online)  Double occupancy $d$ versus density $n$ at fixed inverse temperature $\beta 6J=10$ and different values of the interaction strength $U$ (DMFT).
    Inset: $d(n)/n$ at the inverse temperature $\beta 6J=10$ and different values of the interaction strength $U$ (DMFT).
    }
    \label{fg:D_n_beta10}
\end{figure}

The double occupancy $d = \langle n_{j,\downarrow} n_{j,\uparrow} \rangle$
(Figs.~\ref{fg:D_mu_beta10}, \ref{fg:D_n_beta10})
offers another perspective on the Mott phenomenon.
This quantity is particularly interesting as in cold atom systems it can be directly
measured~\cite{StoeferleEsslinger2006}. Hence, understanding
how double occupancy varies is very useful to characterize the state of an experimental system.
Furthermore, this quantity is directly related to the potential energy of the system which is given for a site by $Ud$.
In Fig.~\ref{fg:D_mu_beta10}, we display the behavior of $d$ as a function of
the chemical potential $\mu$. At weak coupling, $d$ tracks the behavior of the density
(for the non-interacting system, $d= n^2/4$) and the two quantities contain essentially the same
information. However, for large coupling, $d$ is strongly suppressed when the chemical
potential is lower than the potential energy $U$, and increases steeply in the
region $\mu \gtrsim U$, precisely the region where the density increases above
unity. This qualitative difference between weak and strong coupling is well
apparent on Fig.~\ref{fg:D_n_beta10} where we plot $d$ versus $n$. For $U/6J=0.5$, the
curve is nearly quadratic while for $U/6J=5$ the double occupancy $d$ is close to zero for $n<1$ and
linear for $n>1$. The behavior of $d$ at
strong coupling is to be contrasted with the behavior of the density. For strong coupling, $n$
presents a plateau at $n \sim 1$, and hence can be used to determine the
chemical potentials for which $n$ drops below unity as well as for which $n$
exceeds unity. Instead, $d$ is suppressed in the whole region $\mu \lesssim U$, and
is only sensitive to the point at which $n$ exceeds unity.
This is advantageous when it comes to experimentally detecting the
Mott state\cite{DeLeoParcollet2008,ScarolaTroyer2009}.
In the inset of Fig.~\ref{fg:D_n_beta10}, we also display a
plot of $d/n$ as a function of $n$. 

Another quantity of fundamental importance is the entropy as
preparation of cold atom systems is often done almost adiabatically. Therefore, the entropy of the system,
rather than the temperature, is a conserved quantity which can be used as a
constant to characterize the system evolution. We calculate the entropy by integrating
a fit to the DMFT energy data starting from infinite temperature. The high temperature
regime ($\beta6J<1$) is approximated using a second order high temperature expansion.
\begin{figure}[bt!]
    \includegraphics[width=7cm,clip=true]{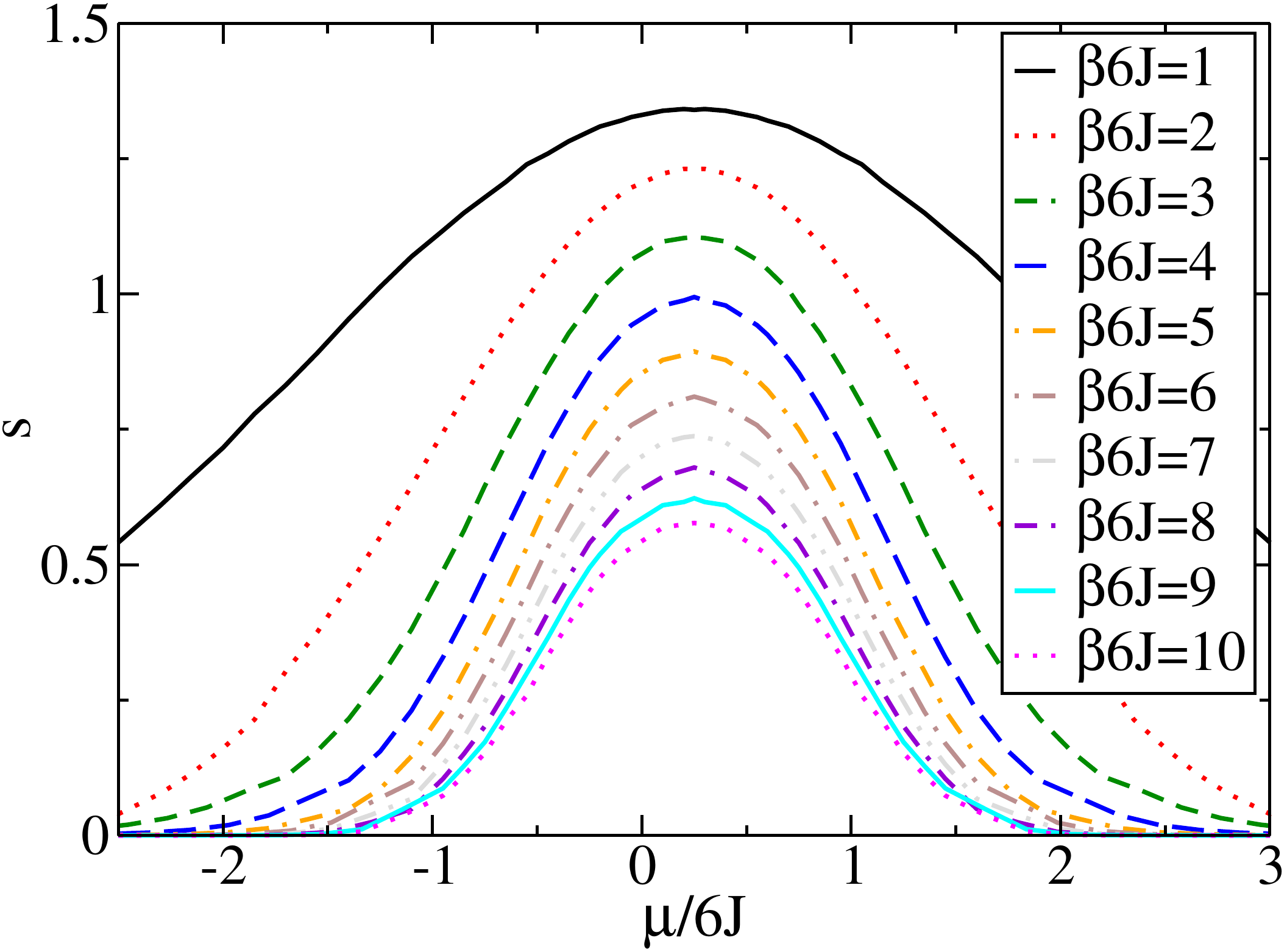}
    \caption{(Color online) Entropy per site $s$ versus chemical potential $\mu$
    at the weak interaction strength $U/6J=0.5$ for different values of the inverse temperature $\beta$ (DMFT).
    }
    \label{fg:s_mu_U0.5}
\end{figure}

In the weakly interacting regime (Fig.~\ref{fg:s_mu_U0.5}), the entropy per site $s$ has a
simple evolution with temperature and chemical potential: it is maximal at $\mu
= U/2$ and decreases monotonically with increasing $|\mu-U/2|$. With decreasing
temperatures, $s(\mu)$ decreases uniformly over the whole chemical potential
range.

\begin{figure}[bt!]
    \includegraphics[width=7cm,clip=true]{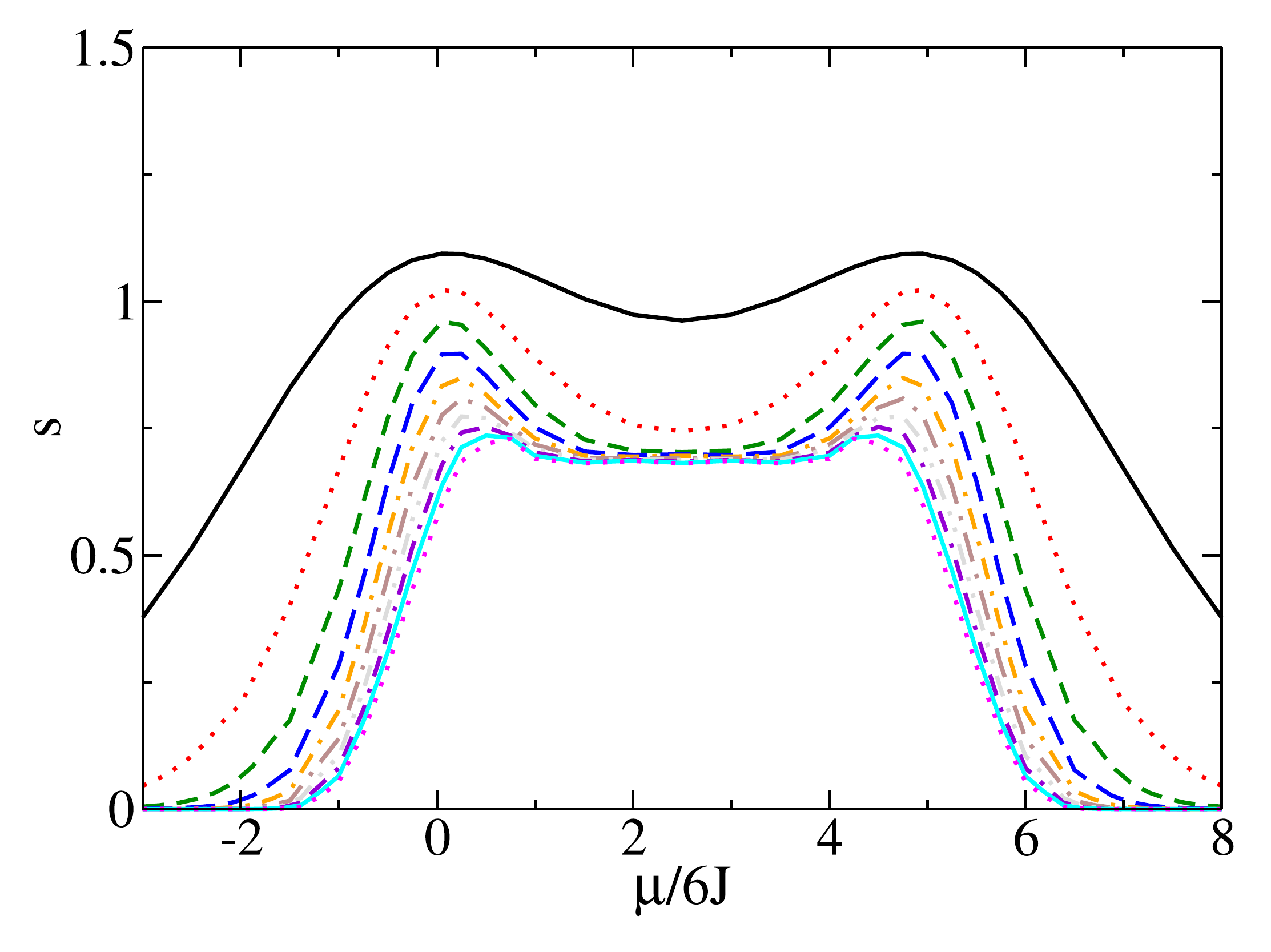}
    \caption{(Color online) Entropy per site $s$ versus chemical
    potential $\mu$ at the strong interaction strength $U/6J=5$ for different
    values of the inverse temperature $\beta$ (DMFT). The curves legend is the same as 
    in Fig.~\ref{fg:s_mu_U0.5}.
    }
    \label{fg:s_mu_U5}
\end{figure}

In the strongly correlated regime (Fig.~\ref{fg:s_mu_U5}), $s(\mu)$ presents
non-trivial features. Already at relatively high temperatures, the entropy has
two maxima located approximately at $\mu=0$ and $\mu=U$, while it has a
minimum at the particle-hole symmetric point $\mu=U/2$.
From an atomic limit point of view, the presence of these two peaks in
the entropy at $\mu=0$ and $\mu=U$ is easily interpreted. These two
chemical potential values correspond to the regions where the density crosses
over rapidly between $n \sim 0$ and $n \sim 1$ ($\mu \sim 0$), and between $n
\sim 1$ and $n \sim 2$ ($\mu \sim U$). The charge fluctuations in these regions are
maximal and contribute significantly to the entropy.

The temperature evolution is also interesting (Fig.~\ref{fg:s_mu_U5}):
the entropy for $\mu \lesssim 0$
and $\mu \gtrsim U$ decreases quickly with lowering $T$, while in the region
$\mu \sim U/2$ the temperature-dependence slows down and the entropy
approaches a finite value, $s(\mu=U/2)\rightarrow\ln 2$.
This value reflects the fact that, in the Mott insulator regime, the system gets frozen
into a local moment state (with two spin-states per site) in a rather extended
temperature range below the Mott gap. Naturally, we expect that, as the system
is cooled down further, the entropy will eventually decrease again below $\ln 2$
as magnetic correlations develop between local moments.
As DMFT neglects spatial correlations in the paramagnetic phase, within this theory,
the decrease in entropy only happens right at the N\'eel transition where
long-range antiferromagnetic order sets in (this N\'eel transition occurs
at a lower temperature than the ones studied in the present paper). However, in reality,
the entropy will start deviating significantly from $\ln 2$ above the N\'eel transition, as short-range
correlations develop~\cite{DareTremblay2007,KoetsierStoof2008,Wessel2010,JoerdensTroyer2010}.
Nevertheless, the single-site DMFT description can be regarded as a good description of the
paramagnetic Mott insulator down to a characteristic temperature. We will see in
Sec.~\ref{sec:theory_validity} how it is possible to evaluate this temperature with the
help of high temperature series expansions.

\begin{figure} [bt]
    \centerline{
    \includegraphics[width=7cm,clip=true]{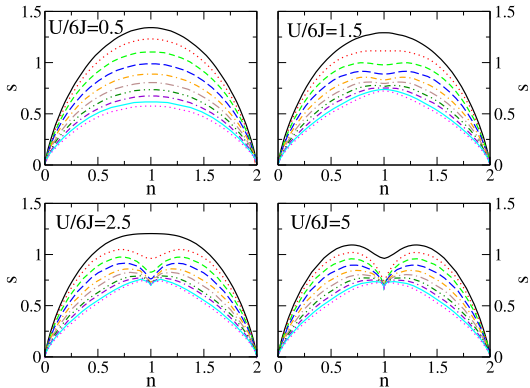}
    }
    \caption{(Color online) Entropy per site $s$ versus density $n$ at different interaction
    strengths $U$ and inverse temperatures $\beta$ (DMFT). The curves legend is the same as 
    in Fig.~\ref{fg:s_mu_U0.5}.
    }
    \label{fig:S_n_hom}
\end{figure}

In Fig.~\ref{fig:S_n_hom}, a similar behavior is observed by plotting the entropy versus the density.
For low interaction strength, a maximum of the entropy is found at half filling.
In contrast, at large interaction strength and intermediate temperatures, a dip in the entropy arises
at half filling, which is due to the freezing of the density degree of freedom in the Mott insulator.
At low temperatures, the entropy decreases linearly in temperature for the liquid
regimes away from half-filling, while the value at half filling remains fixed to $\ln 2$ (for a range
of temperatures, see above).

\begin{figure} [bt]
    \centerline{
    \includegraphics[width=7cm,clip=true]{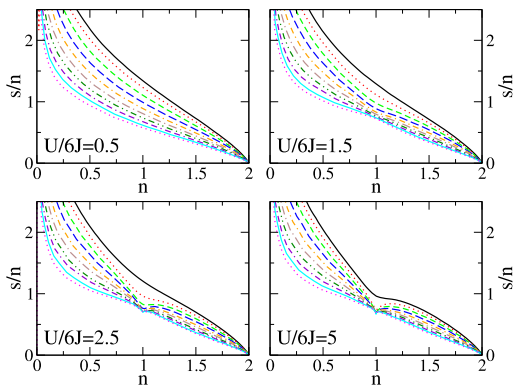}
    }
    \caption{(Color online) Entropy per particle $s(n)/n$ at different interaction
    strengths $U$ and inverse temperatures $\beta$ (DMFT). The curves legend is the same as 
    in Fig.~\ref{fg:s_mu_U0.5}.
    }
    \label{fig:S_over_n_n_hom}
\end{figure}

In Fig.~\ref{fig:S_over_n_n_hom}, we additionally plot the entropy per particle $s/n$
as this quantity will be useful later on to better understand the behavior of the entropy
in the presence of a trapping potential. We note in particular that upon dividing the entropy by the particle number,
the entropy per particle decreases with increasing particle number.

\subsection{Pomeranchuk effect}
\label{sec:pom}

\begin{figure} [bt]
    \centerline{
        \includegraphics[width=7cm,clip=true]{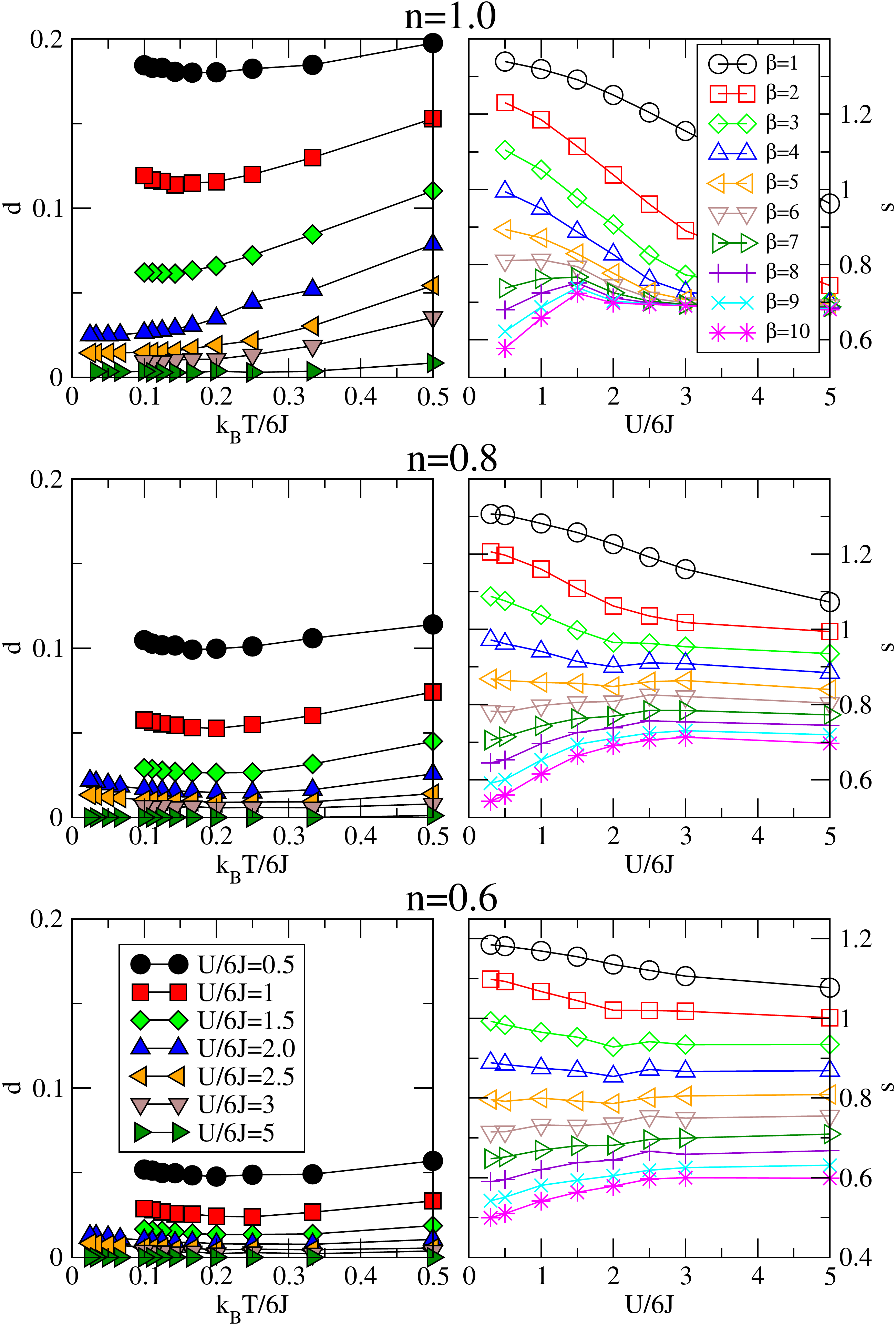}
    }
    \caption{(Color online) Left column: double occupancy $d$ as a function of
    temperature $T$ for different interactions strength $U$ (DMFT). (from top to bottom: $n=1$,
    $n=0.8$ and $n=0.6$). Right column: entropy $s$ as a function of $U$ for
    different temperatures (DMFT). (from top to bottom: $n=1$, $n=0.8$ and $n=0.6$). 
    The inverse temperature $\beta$ is measured in units of $1/6J$.
       \label{fig:pomeranchuk}}
\end{figure}

The double occupancy provides a wealth of information on the physics of
cold atom systems. For example, this quantity can be used as a thermometer
in a certain regime of temperatures and 
interactions\cite{StoeferleEsslinger2006,Koehl2006,DeLeoParcollet2008,JoerdensTroyer2010,KatzgraberTroyer2006}. 
To obtain a good understanding of how this quantity behaves in a trapped system, we first show here the dependence
of the double occupancy on temperature in an homogeneous system (Fig.~\ref{fig:pomeranchuk}).

For moderate interaction strength, the double occupancy presents a
non-monotonous behavior at low temperatures. The appearance of the initial
decrease in the double occupancy with increasing temperature (from $T=0$)
is analogous to the Pomeranchuk effect observed in liquid Helium-3 and has
been discussed previously in the half filled case~\cite{georges_krauth_mott_prb,WernerHassan2005,DareTremblay2007}.
The reason behind this decrease of the double occupancy with increasing temperature
is that the system prefers, when heated, to localize the atoms. For this localized state, the (spin)
entropy is larger than for a state where fermions form a Fermi liquid.
In this regime, the minimum of $d$ is determined by the quasi-particle coherence temperature.
Since, in the particle-hole symmetric case, the coherence temperature decreases with $U$, the
`Pomeranchuk' effect occurs if $U/6J$ is not too large and disappears in the Mott insulator.
Away from particle-hole symmetry, the system behavior is quite different. In that case,
the `Pomeranchuk' temperature at which $d(T)$ has a minimum increases with $U$.
We also note that the double occupancy has recently been studied below and close to the N\'eel
transition to the antiferromagnetic order\cite{GorelikBlumer2010}, a regime that we do not
consider here. In the antiferromagnetic phase, the coherent alignment of spins
causes an increase of the double occupancy.

As the entropy and double occupancy are related by the thermodynamic
(Maxwell) relation~\cite{WernerHassan2005}
\begin{equation}
\frac{\partial s}{\partial U} = -\frac{\partial d}{\partial T},
\label{eq:Max}
\end{equation}
the Pomeranchuk minimum in $d(T)$ translates into a non-monotonous behavior of the
entropy as a function of the interaction strength. For this quantity, a maximum
is found at sufficiently low temperatures. As in the case of the double
occupancy, this effect persists away from half filling (Fig.\ref{fig:pomeranchuk}).
We will see later in Sec.~\ref{sec:isentropic_trap} the consequences of these considerations for
systems subjected to a trapping potential.

In the strong interaction limit, the low-energy properties become similar to the
properties of the Heisenberg model. The entropy for this model has been recently studied in
Ref.~\onlinecite{Wessel2010}.

\subsection{Regimes of validity of DMFT and of high-temperature
series:  a comparative study}
\label{sec:theory_validity}

In this section, we perform a comparative study of the validity of the
two theoretical approximations used in the present paper: DMFT and
high-temperature series. The motivation for doing this is twofold.
On the one hand, in regimes where a converged result can be reliably extracted from
the series expansion (which is an exact technique implying no further approximation),
we can use this comparison to benchmark the validity of DMFT as an approximation
to the thermodynamic properties of the three-dimensional Hubbard model.
On the other hand, by comparing different orders
of the series expansion, we can delineate the regime in which this method can be used
reliably and in which regime DMFT is a better option.

\begin{figure} [bt]
    \includegraphics[width=7cm,clip=true]{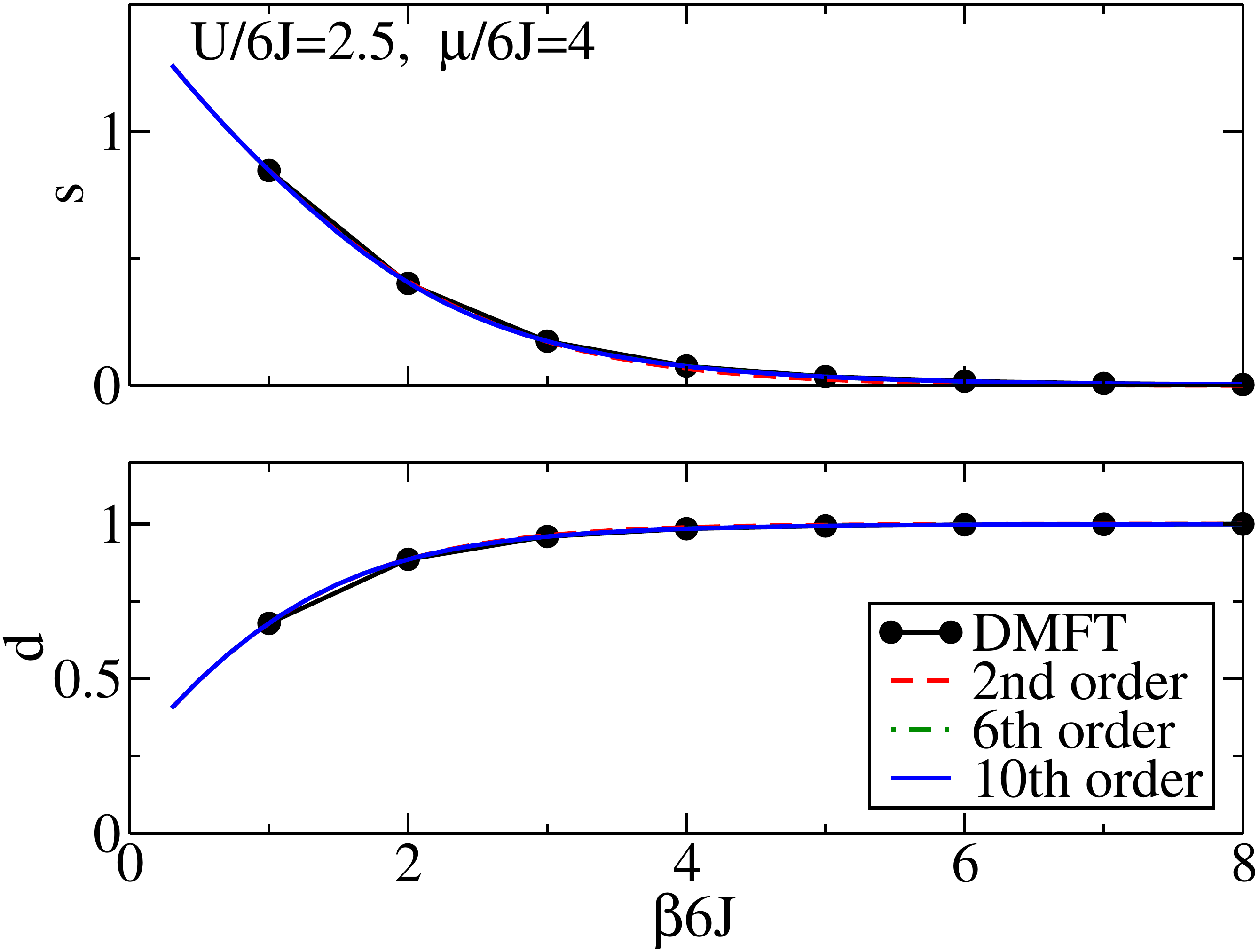}
    \caption{(Color online) Comparison of the results for temperature
    dependence of the entropy $s$ and double occupancy $d$ at fixed
    chemical potential $\mu/6J=4$ and intermediate interaction strength
    $U/6J=2.5$ obtained by high temperature series and DMFT.
    }
    \label{fig:dvsbeta_high_mu}
\end{figure}

\begin{figure} [bt]
    \includegraphics[width=7cm,clip=true]{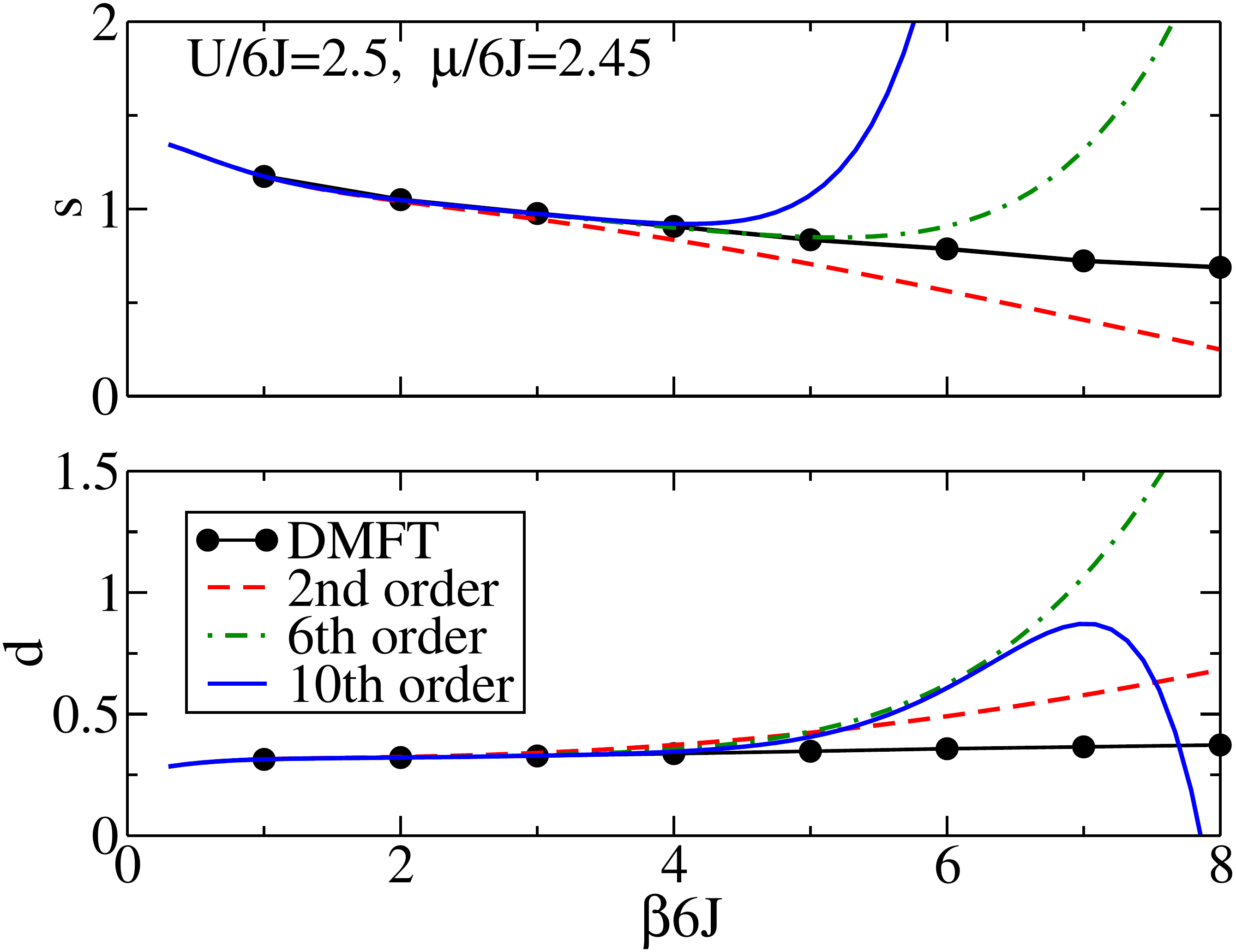}
    \caption{(Color online) Comparison of the results for temperature
    dependence of the entropy $s$ and double occupancy $d$ at fixed
    chemical potential $\mu/6J=2.45$ and intermediate interaction strength
    $U/6J=2.5$ obtained by high temperature series and DMFT method.
    }
    \label{fig:dvsbeta}
\end{figure}

In Fig.~\ref{fig:dvsbeta_high_mu} and Fig.~\ref{fig:dvsbeta}, we display the results obtained by the
two methods for the entropy and the double occupancy as a function of temperature,
at intermediate interaction strength, for two different chemical potentials.
Different orders (up to order $10$) of the series expansions are also compared in these figures.
It is immediately clear from these two figures that the temperature down to which
the series expansion can be safely used, and consequently down to which a reliable
assessment of DMFT can be made, depends considerably on the value of the
chemical potential (or density).
For $\mu/6J = 4$ (Fig.~\ref{fig:dvsbeta_high_mu}), corresponding to a quite high density,
the agreement between the different orders of the series expansion, and the overall
agreement with DMFT is essentially perfect over the whole range of temperatures
considered. In fact, only very small deviations of the second order expansion can actually be detected.
This agreement provides strong evidence for the correctness of both methods down to quite low temperatures
in the high-density regime (and by particle hole-symmetry, also in the low density regime).
In contrast for $\mu/6J=2.45$ (corresponding to an intermediate filling $n\approx 1.3$), it
is clear from Fig.~\ref{fig:dvsbeta} that the different orders of the series expansion start
to deviate from one another already at a rather high temperature $\beta 6J\simeq 4$.
Below this temperature, the series expansion method becomes unreliable.
This breakdown happens approximately at the same temperatures for
the entropy and the double occupancy. As shown below, this regime of intermediate densities is
the hardest one for the series expansion method.

\begin{figure} [bt]
    \includegraphics[width=7cm,clip=true]{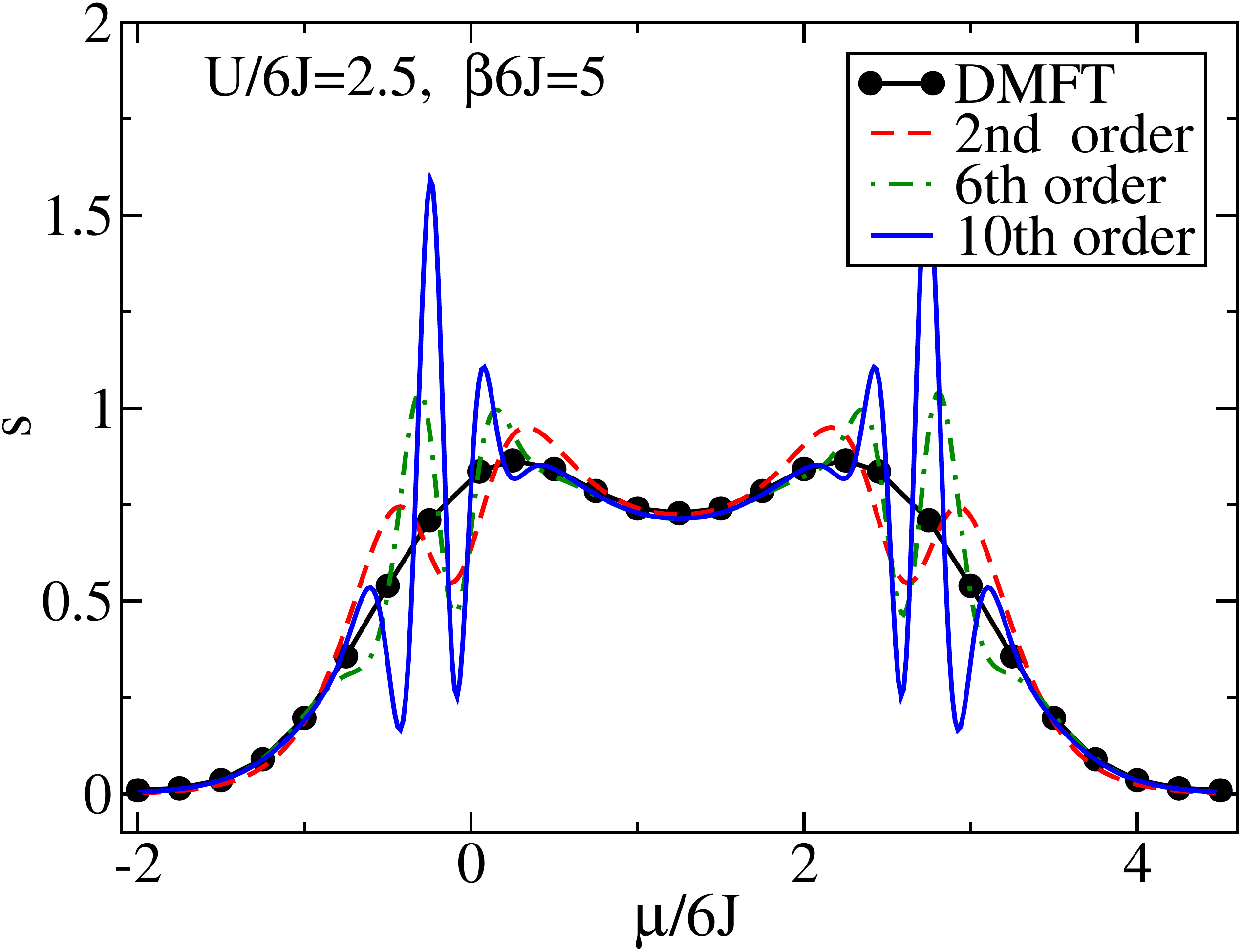}

    \caption{(Color online) Comparison of the results for the entropy $s$
    versus the chemical potential at intermediate interaction strength
    $U/6J=2.5$ obtained by high temperature series and DMFT method.
    }
    \label{fig:svsmu}
\end{figure}

By comparing the different quantities as a function of the chemical potential
at fixed temperature, we identify more clearly these different regimes of density.
In Fig.~\ref{fig:svsmu}, we show the results for the entropy per site $s$ as a function of chemical potential at a
fixed temperature $\beta 6J =5$. Here one can clearly identify three different
regions. At very low (and very high, due to particle-hole symmetry) chemical potential $\mu$,
where the density of the system is close to zero (close to two particles per site), both DMFT and the series
expansion give reliable results down to fairly low temperatures (at least of the
order of $\beta 6J\sim 10$). In contrast, in the intermediate region around $\mu \sim 0$ and
$\mu \sim U$ where the density crosses over from 0 to 1 and from 1 to 2
respectively, the series expansion breaks down at a rather high temperature. Already
at $\beta 6J \sim 5$ it is not reliable anymore.
Very close to half filling, the opposite trend is found. In this region, the series
expansion gives reliable results down to quite low temperatures, actually lower
(as further detailed below) than the temperature at which DMFT ceases to be
an accurate approximation\cite{KozikTroyer2010}.

\begin{figure} [bt]
    \includegraphics[width=7cm,clip=true]{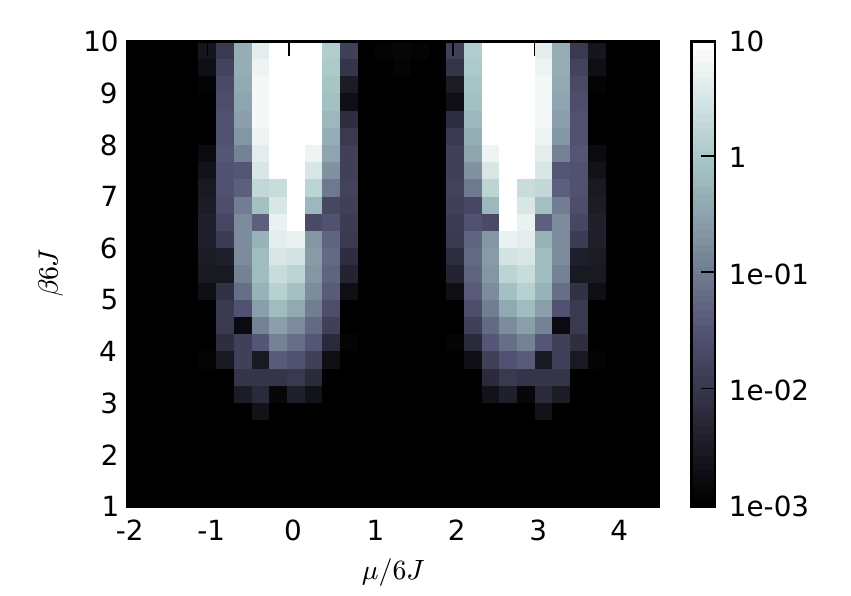}

    \caption{(Color online) Contour plot of the absolute value of the
    difference between the results obtained with
    10th and 6th order series for the entropy versus chemical potential and inverse
    temperature at $U/6J=2.5$.
    }
    \label{fig:diff_series_series}
\end{figure}

In Fig.~\ref{fig:diff_series_series}, we summarize the validity of the high temperature
series expansion by showing the difference between different orders. For this plot, it
appears that the breakdown temperature for the series expansion depends sensitively on the chemical
potential and that the method works best close to very low and very high filling, as well as
exactly at half filling.


\begin{figure} [bt]
    \includegraphics[width=7cm,clip=true]{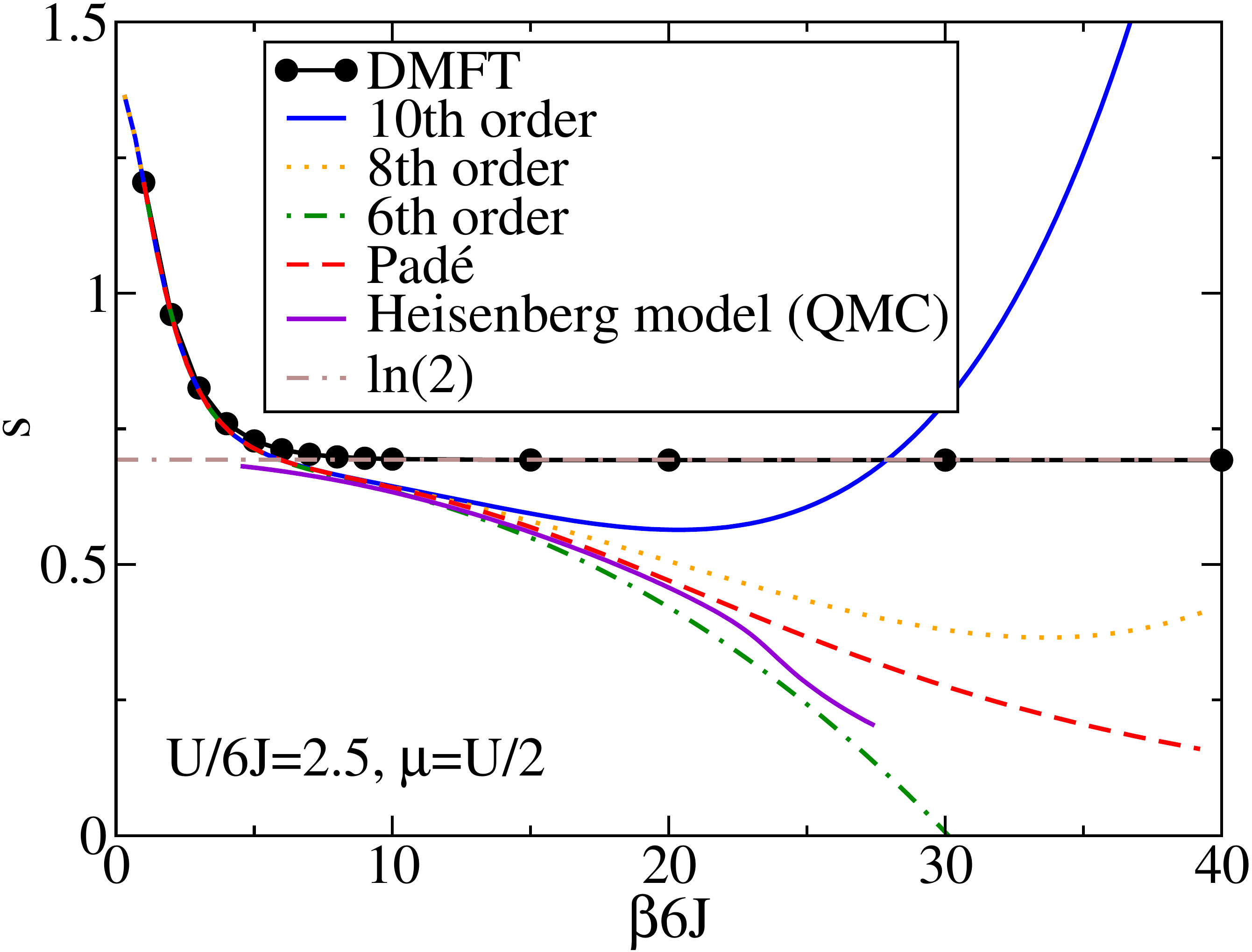}

    \caption{(Color online) Entropy per particle in a system at half filling
    and intermediate interaction strength $U/6J=2.5$ obtained by series
    expansion, DMFT and QMC (for the Heisenberg model)\cite{Wessel2010}.
    \label{fig:svsbeta_pade}}
\end{figure}

In Fig.~\ref{fig:svsbeta_pade}, we focus in more detail on the comparison
of the entropy at half filling ($n=1,\mu = U/2$).
As mentioned previously, the single site DMFT approximation overestimates the
entropy at half filling, yielding a value $\ln 2$ in the large-$U$ limit
because it neglects short-range magnetic correlations.
This value also corresponds to the mean field
value for the entropy of the Heisenberg model in its paramagnetic phase. In reality,
the value of the entropy is reduced by short-range antiferromagnetic correlations\cite{Wessel2010}.
In Fig.~\ref{fig:svsbeta_pade}, this limitation of single-site DMFT is exposed
as we compare entropy obtained with DMFT to the ones given by the series expansion and
the Heisenberg model.
While the DMFT curve saturates at $\ln 2$ at low temperature, the different
orders of the series expansion reach a lower value of the entropy before
diverging. The Pade expansion of the high temperature series expansion lies
slightly above the entropy of the Heisenberg model calculated in Ref.~\onlinecite{Wessel2010} by
QMC simulations. This situation represents a case in which
the high temperature expansion can be used down to a lower temperature
than single-site DMFT.
Finally, we note that further extensions of DMFT
(especially cluster-DMFT methods\cite{Maier_review_2005, kotliar_review2_2006})
exist which overcome these limitations of single-site DMFT, and restore the
physical effects of short-range magnetic correlations.

\section{Trapped system}
\label{sec:trap}

In present day experimental setups, an external potential is usually present to
confine the atomic cloud. This confinement can be due to different sources,
for example to the focusing of the lattice laser beams or a magnetic or dipole trap.
In most cases, a parabolic form is a good approximation for the shape of the confining
potential at the location of the atom cloud. However, many refinements can
be considered to obtain a more precise spatial dependence of the
confining potential $V({\bf r})$ (last term of Eq.~\ref{eq:Hubb_Ham}).

\subsection{Local density approximation and local occupancy histogram}

Within the local density approximation (LDA), the properties of the trapped system at a
certain position $r_j$ are assumed to be those of the homogeneous system with
the chemical potential set to the value $\mu(r_j)$.
LDA has been found to be a good approximation for local quantities
in a three dimensional fermionic gas in an optical
lattice\cite{HelmesRosch2008,GorelikBlumer2010},
and was also validated in the high temperature
regime applicable to ongoing experiments \cite{ScarolaTroyer2009, JoerdensTroyer2010}.
However, as LDA neglects the influence of surrounding sites with different densities,
the main inaccuracies occur close to phase boundaries, where proximity effects occur,
and also when one computes non-local longer range physical observables\cite{KollathZwerger2003,HelmesRosch2008}.

Using LDA and taking the continuum limit, it is possible to describe the system
using rescaled variables that do not depend explicitly on the strength
of the confining potential. To illustrate this we consider a
radial symmetrical form for the trapping potential $V({\bf r})=V_t (|r|/a)^\alpha$ in
$d$ dimensions ($a$ is the lattice spacing). Thus, the position-dependent chemical potential becomes
\begin{equation} \label{eq:mu_r_general_trap}
    \mu(r_j) = \mu_0 - V_t \left(\frac{|r_j|}{a}\right)^\alpha
\end{equation}
with $r_j$ the $d$-dimensional position vector labeling each lattice site and $\mu_0$ the chemical potential
at the center of the trap.
Within LDA, any local observable $O(r_j)$ in the inhomogeneous system is related to
its homogeneous counterpart by $O(r_j)=O_h(\mu(r_j))$, where $O_h$ denotes the
corresponding quantity in the homogeneous system (in the following we will drop
the label `h').

The number of atoms, which corresponds to the sum of the local occupancy over the whole
system, can be expressed in the following way:
\begin{equation}
N=\sum_j n(j) = \frac{\Omega_{d-1}}{a^d} \int d r \, r^{d-1} n(r)
\end{equation}
with $\Omega_{d-1}$ the surface of a sphere in $d$ dimensions. Changing variables
to an integration over the chemical potential using (\ref{eq:mu_r_general_trap})
finally leads to:
\begin{equation}
\rho \equiv\, N\,\left(\frac{V_t}{6J}\right)^{d/\alpha}\,=
\frac{\Omega_{d-1}}{\alpha} \int_{-\infty}^{\mubar_0} d \mubar
\, (\mubar_0 - \mubar)^{\frac{d}{\alpha}-1} n(\mubar). \label{eq:LDA_n}
\end{equation}
In this expression $\mubar\equiv\mu/6J$ is the dimensionless chemical
potential (similarly $\mubar_0$).
The LDA approximation enters in the assumption that $n$
is a function of $\vec{r}$ only through its dependence on $\mu$. This formula
can be easily generalized to the case for which the strength of the confining
potential is different along the different Cartesian directions (which is usually
the case in experiments) by replacing $V_t$ with a proper averaged quantity
$\bar{V}_t$. For example, in three dimensions $\bar{V}_t \equiv (V_{t,x} V_{t,y}
V_{t,z})^{1/3}$.
Eq.~\ref{eq:LDA_n} shows that the dimensionless combination $\rho=N (V_t/6J)^{d/\alpha}$ is a
quantity that does not depend on the strength of the confining potential, and
hence can be used to describe properties of experimental systems regardless of
the particular realization of the trap.
For the one-dimensional case, see Ref.~\onlinecite{RigolScalettar2003}.
It also shows that in this approximation the key quantity is the observable
in the homogeneous system, e.g., $n(\mu)$, and that everything can be derived from it.
Obviously the same holds true for all the other quantities that can be
expressed as a sum of a local quantity over the whole system.

Another very useful way to express averages of observables over the trap is to introduce the
distribution function of site occupancies in the system, defined for $0\leq n\leq 2$ as:
\begin{equation}
    P(n) \equiv \sum_j \delta(n_j-n),
\end{equation}
and the related quantity in which each site is weighted by its occupancy:
\begin{equation}
    Q(n) =\sum_j n_j \delta(n_j-n) = nP(n).
\end{equation}
Using again a continuous-space integration over the system and changing variables in favor of the
chemical potential, one obtains:
\begin{equation}
    q(n)\equiv (V_t/6J)^{d/\alpha}Q(n)\,=\,
    \frac{\Omega_{d-1}}{\alpha} \frac{n}{\kappa(n)}\,
        \left[\mubar_0 - \mubar(n)\right]^{d/\alpha-1}. \label{eq:q_n_def}
\end{equation}
In this expression, $\kappa(n)$ is the (dimensionless) compressibility of the homogeneous system:
\begin{equation}
    \kappa(n)= \left. \frac{\partial n}{\partial \mubar}\right|_{\mu=\mu(n)}.
    \label{eq:compress}
\end{equation}
Hence, the local chemical potential and local compressibility entirely determine the
distribution of local site occupancies in the trap.
The distribution $Q(n)$ and rescaled distribution $q(n)$ obey the sum-rules:
\begin{equation}
    \int_0^2 dn \, Q(n) = N\,\,\,,\,\,\,
    \int_0^2 dn \, q(n) = \rho
    \label{eq:sumrule}
\end{equation}
with $N$ the total atom number and $\rho=(V_t/6J)^{d/\alpha} N$ its rescaled form.

These distribution functions allow to express the average of any observable
over the trap, within the LDA approximation, as:
\begin{equation}
    O\,\equiv\,\sum_j o(j)\,=\,\int_0^2 dn \, P(n) o(n)=
    \int_0^2 dn \, Q(n) \frac{o(n)}{n}
    \label{eq:av_dist}
\end{equation}
where $o$ is the local operator corresponding to the observable. Because
$Q(n)$ is a normalized distribution obeying the sum-rule (\ref{eq:sumrule}), the
last expression is particularly useful. As we shall see later, when varying an
external parameter, it allows to
separate in a simple manner the changes in $O$ which are due to a redistribution of
the particles in the trap (reshaping of $Q(n)$) from the contribution due to the intrinsic
dependence of the local observable on the parameter, already present in the
homogeneous system.

In the following, we make use of this description, in combination
with thermodynamic relations, in order to discuss cooling or heating of
the trapped system as the coupling is changed. We will concentrate on the
case of a three-dimensional lattice in a harmonic potential
i.e., $d=3$, $\alpha=2$.

\subsection{State diagram}

One consequence of the presence of an inhomogeneous trapping potential is that
different quantum phases can spatially coexist in the gas.
This can actually be seen as a favorable situation, in which several different
physical regimes can be studied in a single experiment.
In an optical lattice realizing the three-dimensional Hubbard model,
coexistence between liquid and Mott insulating regions in the trap were
for example documented in theoretical studies~\cite{HelmesRosch2008,DeLeoParcollet2008}.

\begin{figure} [bt]
\includegraphics[width=7cm,clip=true]{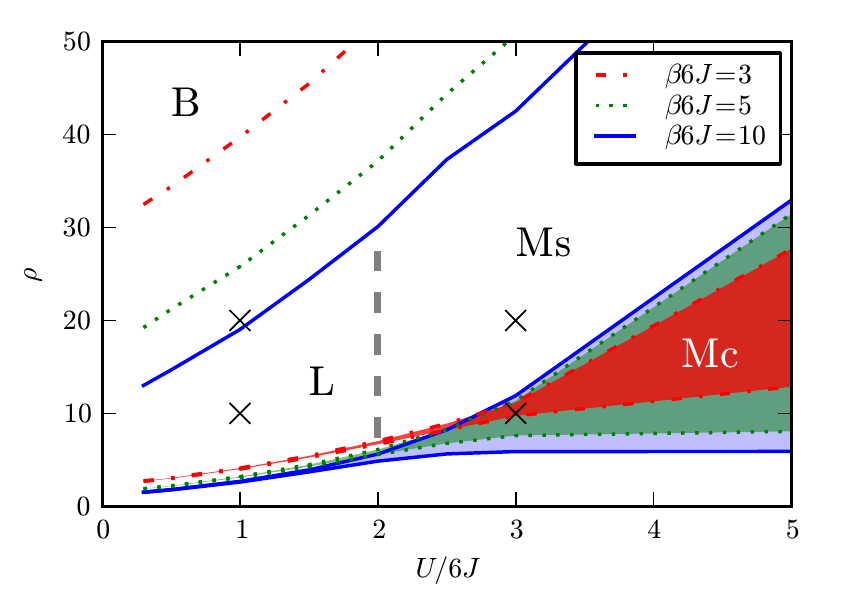}

    \caption{(Color online) State diagram of the gas in a three-dimensional
    optical lattice with parabolic trapping, for different
    temperatures (DMFT). The four characteristic regimes (see text) are labeled
    by: B (band insulator in the center of the trap), Mc (Mott insulator in the
    center of the trap, shaded areas), Ms (shell of Mott insulator away from the center) and
    L (liquid state).
    For each temperature the (crossover) lines indicate, from bottom to top,
    the $\rho$ values at which the central density takes the values 0.995,
    1.005 and 1.995. The gray dashed line marks
    the crossover from the liquid to the Mott state with increasing
    interaction.
    The crosses indicate the points for which the density
    profiles are plotted in Fig.~\ref{fig:profiles}.
    }
    \label{fig:state_diagram}
\end{figure}

In Fig.~\ref{fig:state_diagram}, we display the different regimes expected in
a three-dimensional optical lattice confined into a parabolic trap, as a function of the
coupling $U/6J$ and of the scaled particle number $\rho$. Different temperatures in the
currently accessible range are considered. At still lower temperature (not displayed),
antiferromagnetic long-range order~\cite{SnoekHofstetter2008} will occur in the regimes
with a commensurate Mott plateau.
The state diagram of Fig.~\ref{fig:state_diagram}, which generalizes to different temperatures the
results of Ref.~\onlinecite{DeLeoParcollet2008}, was obtained on the basis of the
theoretical calculations described in the previous section for the homogeneous model, using LDA approximation.
In addition, in Appendix~\ref{sec:simpleapprox}, we introduce a simple approximation which allows to
obtain analytical expressions for the various crossover lines of the state diagram in
the low-temperature regime.

\begin{figure} [bt]
\centerline{
\includegraphics[width=8.5cm,clip=true]{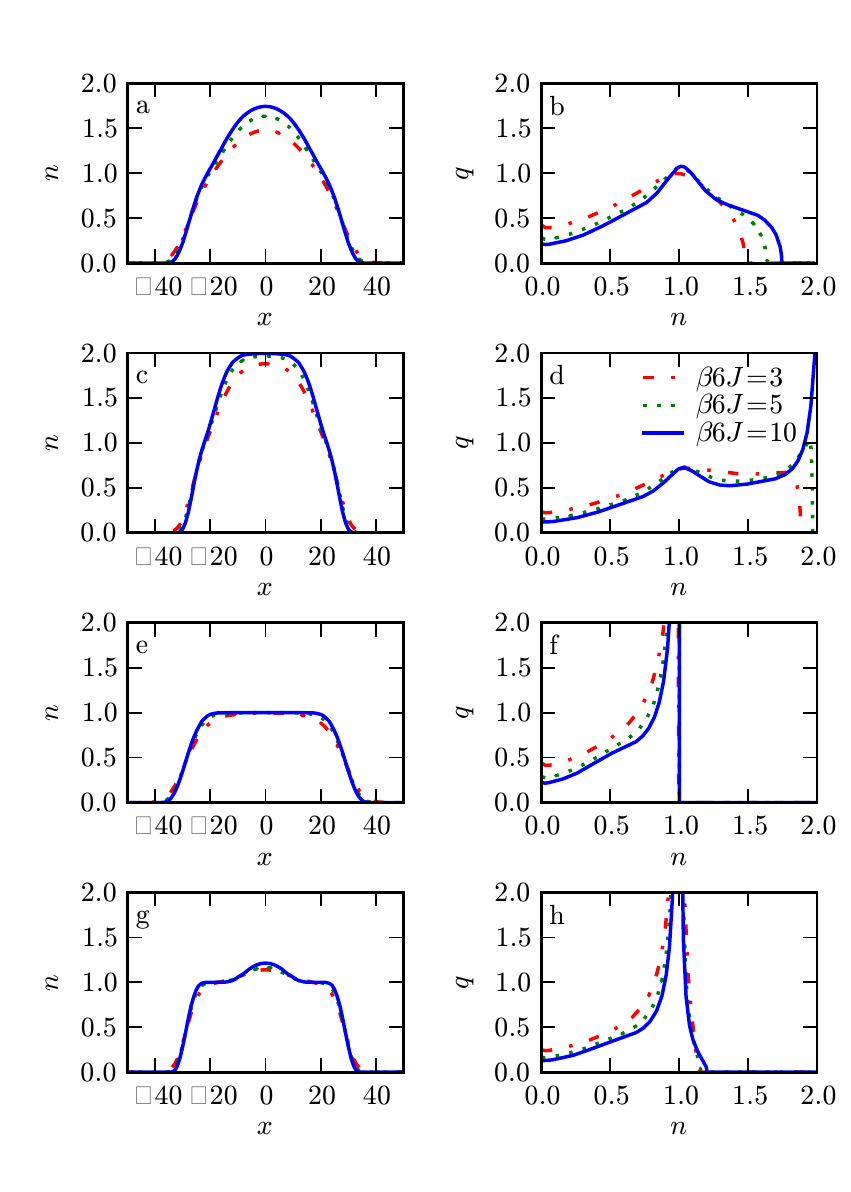}
}

    \caption{(Color online) Density profiles (left column) and occupancy
    distribution q(n) (right column) for
    four typical points in the state diagram (obtained with DMFT):
    a and b (in regime `L'): $U/6J=1$, $\rho=10$.
    c and d (in `B' for low-$T$): $U/6J=1$, $\rho=20$.
    e and f (in `Mc'): $U/6J=3$, $\rho=10$.
    g and h (in `Ms'): $U/6J=3$, $\rho=20$.
}
    \label{fig:profiles}
\end{figure}

The state diagram displays four characteristic regimes (labeled L, B, Mc and Ms),
which are illustrated by the four
corresponding density profiles $n(r)$ and local occupancy distribution functions
$q(n)$ calculated at four representative points and displayed in Fig.~\ref{fig:profiles}.

For low interaction strength (regime `L', Fig.~\ref{fig:profiles} a) the density profile adjusts
to the trapping profile and the system remains a Fermi liquid everywhere in the trap.
With increasing temperature the density distribution broadens. The weighted particle number
distribution $q(n)$ displays a maximum at filling unity which, according to Eq.~(\ref{eq:q_n_def})
reflects the smaller compressibility at that filling.
A rather sharp drop is seen towards larger occupancies which represents
the center of the trap, whereas a slower decay occurs towards smaller particle numbers,
due to the tails of the density distribution.
Increasing the temperature shifts weight from larger occupancies towards smaller occupancies.

For very large values of the scaled particle number $\rho$, a band insulator with $n=2$ forms in the
center of the trap (regime `B', Fig.~\ref{fig:profiles} c). The pinning to
$n=2$, and hence the band insulator, is destroyed by
increasing the temperature. In the presence of a large band insulating region,
the corresponding distribution $q(n)$ displays a sharp peak at filling $n=2$.
Increasing the temperature, this peak decreases and the weight moves to lower occupancies.

For larger interaction strength (regime `Mc', Fig.~\ref{fig:profiles} e) a Mott-insulating region appears,
in which the density is pinned to $n=1$ particle per site.
Close to the boundary of the trap, the Mott insulating region is surrounded by a liquid region.
The Mott-insulating region shows up in $q(n)$ as a large and narrow peak at filling $n=1$ with a sharp edge
on the large occupation numbers side. The peak reflects the essentially vanishing
compressibility of the Mott insulator.
Increasing the temperature decreases the size of the Mott insulating plateau, and results in a shift of
the weight from filling one to low densities.

Increasing the number of atoms in the trap at large interaction strength can increase the
pressure exerted on the atoms, and can cause the occurrence of a liquid region with
filling larger than one in the center, surrounded by a shell of Mott insulator with $n=1$
(regime `Ms', Fig.~\ref{fig:profiles} g).
Correspondingly, the sharp peak in $q(n)$ broadens somewhat.

Recently, experimental evidence of the Mott insulating region
has been reported\cite{JoerdensEsslinger2008,SchneiderRosch2008}.
This has been achieved by observing the suppression of the double occupancy
in the Mott-insulating region\cite{JoerdensEsslinger2008,DeLeoParcollet2008, ScarolaTroyer2009}
and the compression of the cloud as a response to the variation of the external
trapping potential\cite{SchneiderRosch2008}.

We note that in bosonic two-dimensional gases the density profiles $n(r)$
and therefore the occupancy number distributions $q(n)$ can nowadays be measured with a
very high spatial resolution\cite{GemelkeChin2009,BakrGreiner2009,ShersonKuhr2010}.
In three dimensional gases, the integrated
column density can be measured for example by using an electron microscope\cite{WuertzOtt2009}.
Furthermore, new techniques such as the immersion of a single trapped ion into the atomic gas
are being developed to locally measure the density in three dimensional systems as well\cite{ZipkesKoehl2010}.

\subsection{Temperature changes in the trap during
an adiabatic evolution}
\label{sec:isentropic_trap}

Cold atom clouds are almost perfectly isolated from their environment. Therefore, assuming that
manipulations can be performed adiabatically, the quantity that is conserved during the
evolution of the system is the entropy. 
However, the temperature will in general change.
Consequently, studying the effects of an isentropic change of parameters on
the system temperature is important. Here we focus our attention on the increase of the interaction strength, $U$,
and identify if such a change can help to reach lower temperatures in a trapped system, i.e. whether
cooling occurs.

\begin{figure} [bt]
\includegraphics[width=6cm,clip=true]{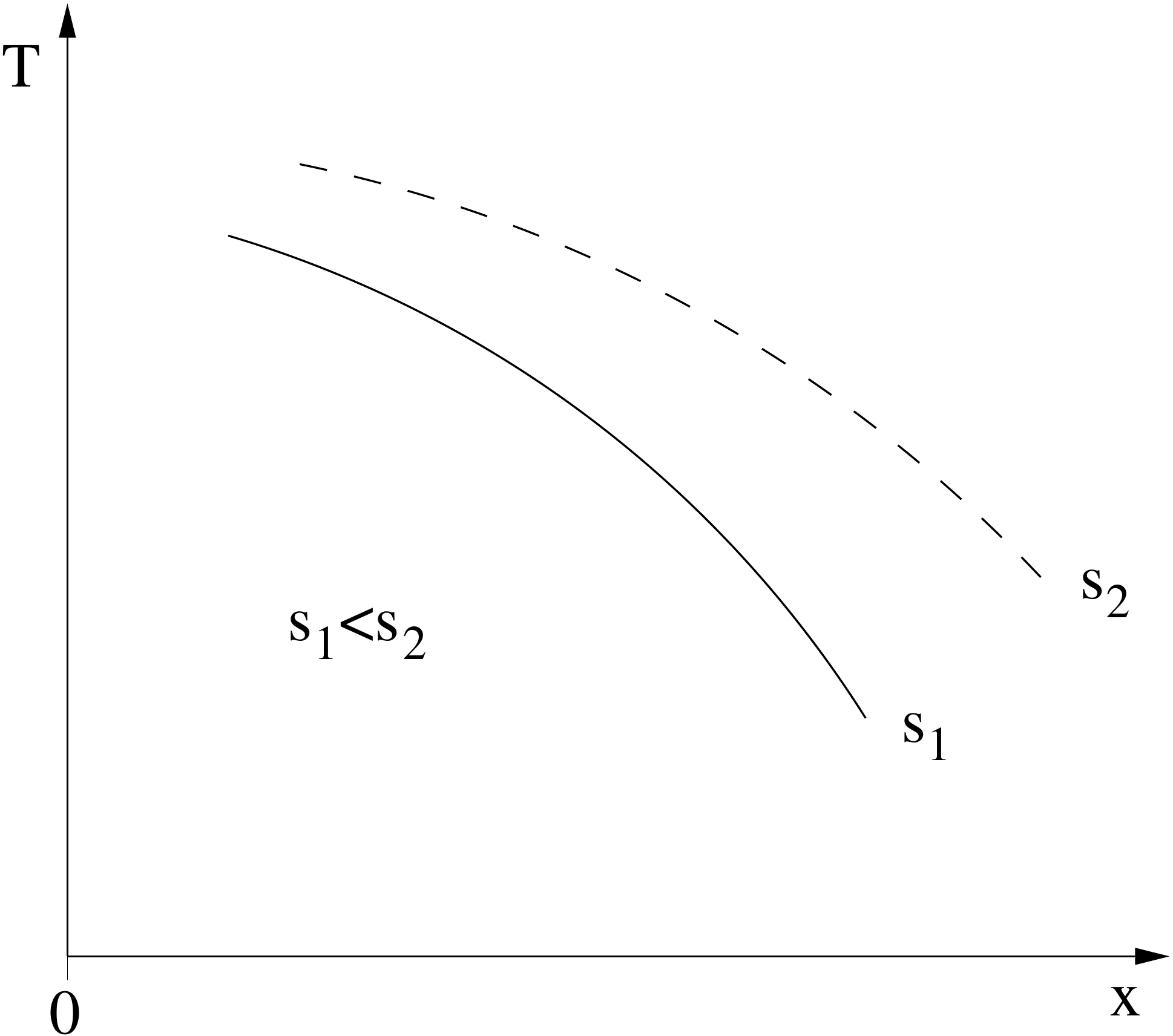}
    \caption{Schematic evolution of the temperature as a function of a
    parameter $x$ along isentropic lines, in a case where
    cooling occurs as $x$ is increased. The continuous line corresponds to a
    lower entropy per particle than the dashed line, $s_1 < s_2$.
    The figure illustrates that for this situation, increasing $x$ while keeping $T$ constant
    will increase the entropy (Eq.~\ref{eq:cooling_rate}).}
    \label{fig:temperature_change}
\end{figure}

The change in temperature induced by a change in a certain parameter $x$ at
constant entropy is $\delta T/ \delta x |_S$. If this quantity is negative (positive), an
increase of the parameter is associated with cooling (heating).
As we now show, one can consider alternatively the change in entropy at constant
temperature $\delta S/ \delta x|_T$. Indeed, the location $T(x)$ of isentropic lines
in the $(x,T)$ plane is defined by the equation $S[T(x);x]=\rm{const.}$. Taking a derivative of
this equation, we obtain:
\begin{equation}
    \left.\frac{\delta T}{\delta x}\right|_S\,
    \left.\frac{\delta S}{\delta T}\right|_x\,+\,
    \left.\frac{\delta S}{\delta x}\right|_T = 0.
\end{equation}
Denoting by $c=T\delta S/\delta T$ the specific heat of the system, we finally obtain the expression
of the relative cooling rate in the form:
\begin{equation}
    \left.\frac{1}{T}\,\frac{\delta T}{\delta x}\right|_S\,=\,
    -\left.\frac{1}{c}\,\frac{\delta S}{\delta x}\right|_T.
    \label{eq:cooling_rate}
\end{equation}
Since $c$ is a positive quantity, we see that there is cooling (heating) when
$\left.\frac{\delta S}{\delta x}\right|_T$ is positive (negative).
This is illustrated schematically in Fig.~\ref{fig:temperature_change}.

In the following, we consider the temperature change under an adiabatic increase of the
coupling $U$. We observe that the derivative $\left.\frac{\delta S}{\delta U}\right|_T$ is related to
a derivative of the total double occupancy $D=\frac{1}{N}\sum_j \aver{n_{\uparrow,j}n_{\downarrow,j}}$
through the Maxwell relation~\cite{WernerHassan2005}:
\begin{equation}
    \label{eq:dDdT}
    \left. \frac{\delta S}{\delta U}\right|_{N,T}=
    \left. -\frac{\delta D}{\delta T}\right|_{N,U}.
\end{equation}
So that the relative cooling rate reads:
\begin{equation}
    \left. \frac{1}{T}\,\frac{\delta T}{\delta U}\right|_{S,N}\,=\,
    \left. \frac{1}{c}\,\frac{\delta D}{\delta T}\right|_{N,U}.
    \label{eq:cooling_rate_U}
\end{equation}
Hence, when the derivative of $D$ with respect to temperature is negative
(positive) there will be cooling (heating) upon an isentropic increase
of the interaction strength.
One advantage to use the change of $D$ is that this quantity can be measured
quite accurately in present experiments\cite{StoeferleEsslinger2006, JoerdensEsslinger2008}.

\begin{figure}[bt!]
    \includegraphics[width=7cm,clip=true]{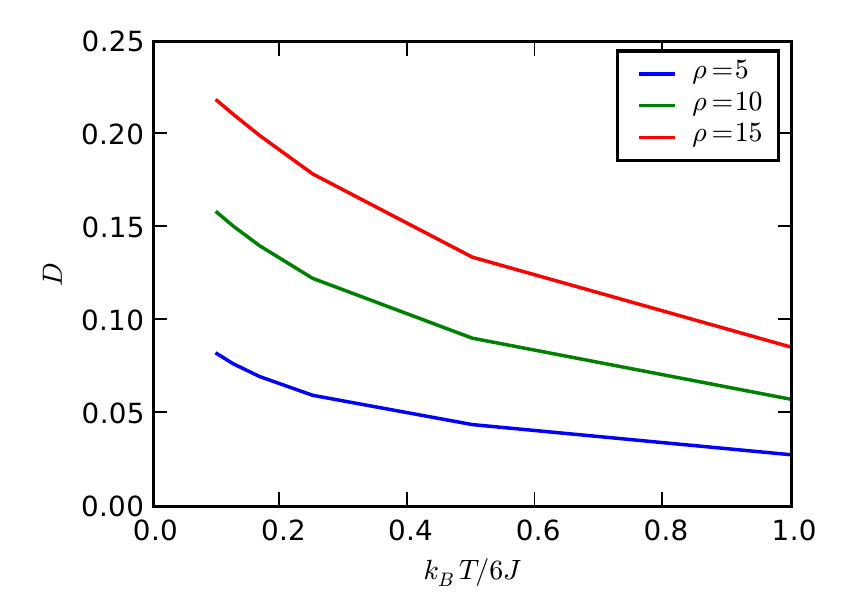}
    \caption{(Color online) Temperature dependence of the number of doubly occupied sites $D(T)$ for the interaction strength $U/6J=1$
             and at different scaled particle numbers $\rho=5,10,15$ (DMFT).
        }
    \label{fg:DvsT_U1}
\end{figure}

\begin{figure}[bt!]
    \includegraphics[width=7cm,clip=true]{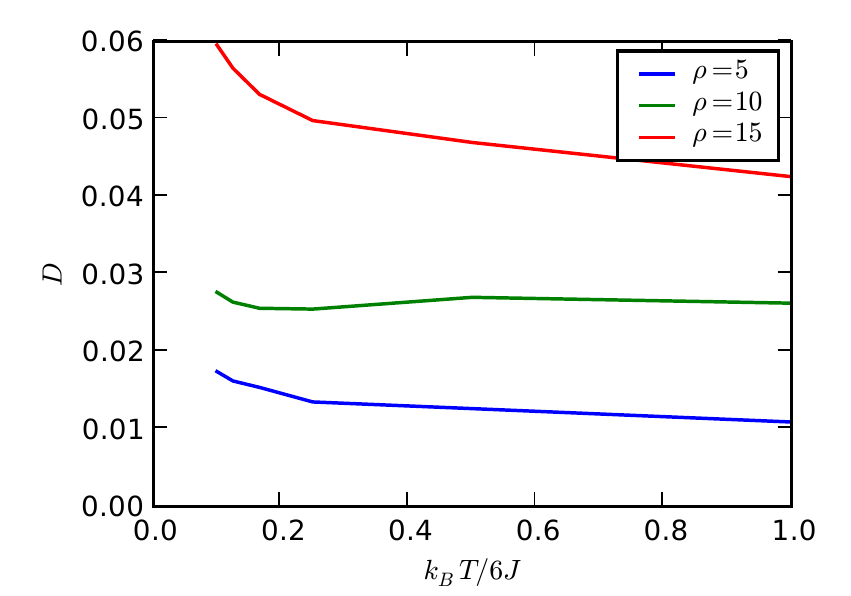}
    \caption{(Color online) Temperature dependence of the number of doubly occupied sites $D(T)$ for the interaction strength $U/6J=2$
             and at different scaled particle numbers $\rho=5,10,15$ (DMFT).
    }
    \label{fg:DvsT_U2}
\end{figure}

\begin{figure}[bt!]
    \includegraphics[width=7cm,clip=true]{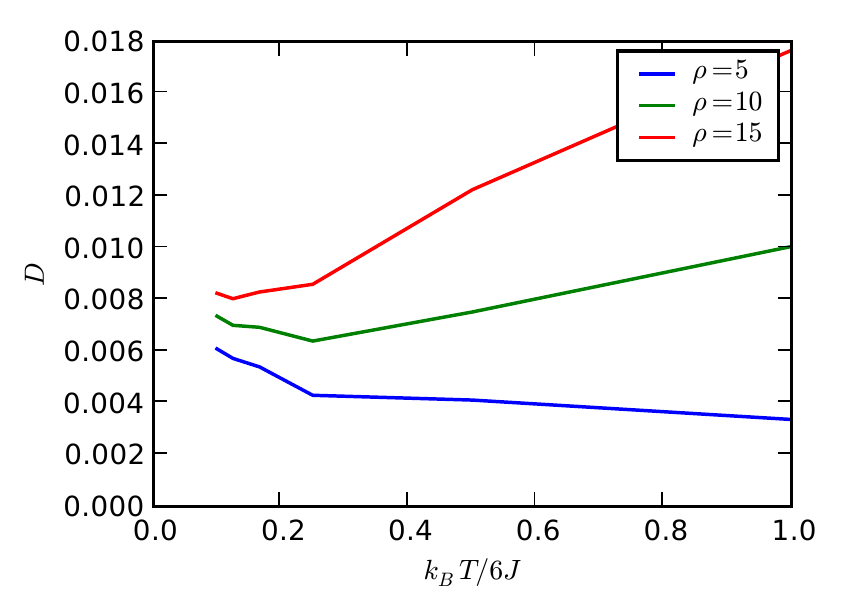}
    \caption{(Color online)  Temperature dependence of the number of doubly occupied sites $D(T)$ for the interaction strength $U/6J=3$
             and at different scaled particle numbers $\rho=5,10,15$ (DMFT).
    }
    \label{fg:DvsT_U3}
\end{figure}

In Fig.~\ref{fg:DvsT_U1}-\ref{fg:DvsT_U3}, we plot $D$ as a function of
temperature for different interaction strengths and particle
numbers. For $U/6J=1$ (Fig.~\ref{fg:DvsT_U1}), $D(T)$ is a decreasing function of temperature
for all particle numbers. This implies that in the weak-coupling regime an
increase of $U$ generates cooling. Increasing the interaction the situation
gradually changes. For $U/6J=2$ (Fig.~\ref{fg:DvsT_U2}), $D(T)$ becomes much flatter and
cooling is restricted to large particle numbers or low temperatures. Finally,
for $U/6J=3$ (Fig.~\ref{fg:DvsT_U3}), the tendency inverts and at high temperature,
heating occurs. It has to be stressed that the absolute value of the
derivative drops with increasing interaction and that at large $U$ the heating or cooling
is essentially negligible.

In order to better understand the origin of the cooling or heating, we write
the derivative $\delta D/\delta T|_{N,U}$ using Eq.~(\ref{eq:q_n_def}), under the form:
\begin{equation} \label{eq:dDdT_qfunc}
    \left. \frac{\delta D}{\delta T}\right|_{N,U} =
    \int_0^2 \frac{dn}{n} \left[ q(n) \partial_T d|_n + d \, \partial_T q(n)
    \right].
\end{equation}
The usefulness of this expression resides in the clear separation of two
different contributions. The first term in the integrand takes into
account changes to the total double occupancy in the trap due to the intrinsic temperature
dependence of the double occupancy $d(n,T,U)$ in the homogeneous system.
Cooling can occur from this term whenever the Pomeranchuk mechanism discussed
above applies.
The second term instead represents the contribution due to the
redistribution of the atoms in the trap upon a temperature change.
The two terms can be calculated separately in order to determine the most
relevant mechanism behind the cooling observed for low interaction.

\begin{figure}[bt!]
    \includegraphics[width=7cm,clip=true]{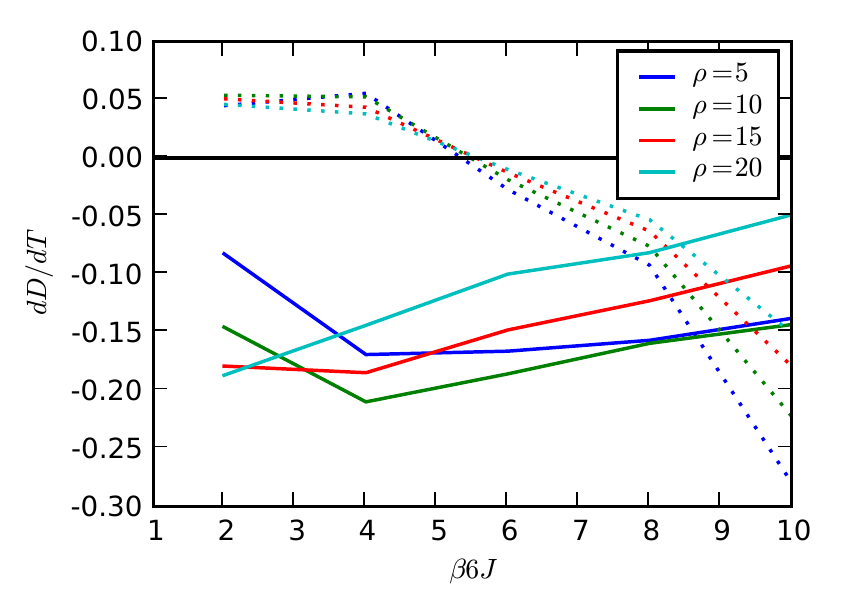}
    \caption{(Color online) The two contributions to $\delta D/\delta T$ for
    $U/6J=1$ (DMFT). Continuous line (left side, top to bottom: 
    $\rho = 5, 10, 15, 20$): redistribution of atoms in the trap (reshaping,
    second term in Eq.~(\ref{eq:dDdT_qfunc})).
    Dotted line (right side, top to bottom: $\rho = 20, 15, 10, 5$): intrinsic 
    change of the double occupancy
    with temperature (first term in Eq.(\ref{eq:dDdT_qfunc})).
    }
    \label{fg:pomer_contributions_lowU}
\end{figure}

In Fig.~\ref{fg:pomer_contributions_lowU}, we plot the first (``intrinsic'',
dotted lines) and second (``redistribution'', continuous lines) terms of
Eq.~\ref{eq:dDdT_qfunc} at $U/6J=1$ for different atom numbers $\rho$.
Note that, for readability, these quantities are plotted versus {\it inverse}
temperature $\beta=1/k_BT$.
We notice that the ``redistribution'' term is always negative, and hence the
reshaping of the density profile always induces a cooling effect.
In contrast, the ``intrinsic'' term is positive at high temperature and becomes negative only
for $\beta 6J \gtrsim 5$ ($k_BT/6J \lesssim 0.2$).
Hence, we conclude that at high temperature the cooling is dominated by the redistribution
of atoms in the trap, while at lower temperatures, both the redistribution and the
intrinsic `Pomeranchuk' effect contribute on comparable footing. The latter may even
dominate at still lower temperatures
(e.g., $k_BT/6J \lesssim 1/8$ in Fig.~\ref{fg:pomer_contributions_lowU}).

A qualitative understanding of the behavior of each term in Eq.~(\ref{eq:dDdT_qfunc})
can be achieved from the inspection of the properties of $d(n)/n$ and $q(n)$.
The first observation is that $d(n)/n$ is a monotonically increasing function
of $n$ (cf.~inset of Fig.~\ref{fg:D_n_beta10} ). Secondly, from Fig.~\ref{fig:profiles}b and d
we notice that $\partial_T q(n)$ (possibly
with the exception of the region around $n=1$) is always negative for $n$
larger than a certain value $\bar{n}$ and always positive for $n$ smaller than
$\bar{n}$. Furthermore $\int dn \, q(n) = \rho$ implies that $\int dn \,
\partial_T q(n) = 0$. Combining these two observations, we conclude that the
second term in Eq.(\ref{eq:dDdT_qfunc}) is generally negative. Hence, the
redistribution of the atoms indeed produces cooling in general, as observed above.
The presence of the peak in $q(n)$ around $n=1$ might undermine the reasoning but
this is never the case for the parameters considered here (Fig.~\ref{fg:pomer_contributions_lowU}).

The behavior of the first term in Eq.~\ref{eq:dDdT_qfunc} is closely connected
to the Pomeranchuk effect in homogeneous systems. As we saw in Fig.~\ref{fig:pomeranchuk}
in the homogeneous system the Pomeranchuk effect is active only at sufficiently
low temperature. Therefore, the intrinsic contribution in the trap can only
lead to cooling at low temperatures. This is in good agreement with
the results in Fig.~\ref{fg:pomer_contributions_lowU}.

\begin{figure}[bt!]
    \includegraphics[width=7cm,clip=true]{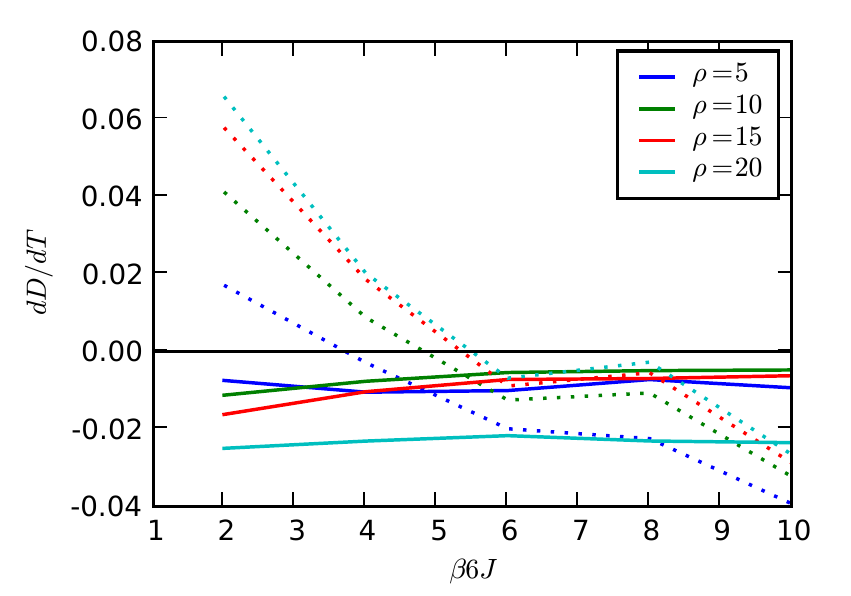}
    \caption{(Color online) Different contributions of the intrinsic and
    the reshaping term to $\delta D/\delta T$ for $U/6J=3$ (DMFT). Continuous line (left side,
    top to bottom: $\rho = 5, 10, 15, 20$) is the reshaping,
    dotted (left side, top to bottom: $\rho = 20, 15, 10, 5$) is 
    the intrinsic double occupancy change with temperature.
    }
    \label{fg:pomer_contributions_highU}
\end{figure}

The situation is only quantitatively different at stronger coupling. For $U/6J=3$
(Fig.~\ref{fg:pomer_contributions_highU}), the redistribution of atoms still corresponds
to cooling, although the absolute value of the contribution is
roughly an order of magnitude smaller than for weaker interaction strength.
However, in this case, the `intrinsic' contribution becomes dominant and its sign
corresponds to heating (opposite to the Pomeranchuk effect).
At lower temperature, the intrinsic term becomes negative again (cooling) but anyhow
the cooling rate in this regime is fairly small.

The conclusion of this section is that in trapped systems an increase of the
interaction is accompanied by cooling for interactions weaker than the
interaction needed to have a Mott insulator. On the other hand, at larger
interactions, there is no substantial cooling associated with an increase of $U$,
and even slight heating can occur at high temperature. On the whole,
the dominant contribution to cooling is usually the redistribution of the
atoms in the trap, although the Pomeranchuk effect (intrinsic contribution)
can become operative at low temperature.

\subsection{Cooling by trap shaping}

Reaching sufficiently low temperatures to observe complex quantum phases is one of
the main challenges currently faced by cold atom physics experimentalists. 
In this section, we show that by adiabatically reshaping the confining trap, to divide the
system into entropy rich and poor regions, a gas can be cooled down by one order of magnitude lower than 
currently achievable using state-of-the-art techniques~\cite{BernierKoehl2009}. 
Before presenting our method, we would like to point out that employing adiabatic 
changes to cool down gases was proposed in related contexts. 
For example, a Bose-Einstein condensate was experimentally produced by adiabatically deforming 
the external trapping potential to increase the gas phase space 
density~\cite{PinkseWalraven1997,StamperKurnKetterle1998}. Furthermore, reshaping the underlying trap to
create entropy rich regions that are later isolated from the remaining system was proposed for 
bosons loaded into an optical lattice~\cite{PoppCirac2006,CapogrossoSvistunov2008}. 
For fermions confined to an optical lattice, cooling could be achieved by immersion 
into a bosonic bath~\cite{GriessnerZoller2006,HoZhou2009}. In Ref.~\onlinecite{HoZhou2009}, it 
was cleverly proposed to reduce the entropy of lattice fermions in contact with a bosonic reservoir
by compressing them into a band insulator or more generally a gapped phase. This last cooling method requires 
a transfer of entropy between two distinguishable quantum gases, a process experimentally demonstrated 
in Ref.~\onlinecite{CataniInguscio2009}. 

Our cooling method does not rely on immersing the atoms into a reservoir but on creating spatially 
distinct regions of high and low entropy that can be isolated from one another, and on subsequently
removing the high entropy region. Once this procedure is completed the remaining system has a much 
lower average entropy per particle allowing for the study of interesting phenomena requiring lower 
temperatures than previously attainable. To demonstrate our cooling procedure, we use a twofold approach. 
We first present our method using an idealized setup, and in a second time, we revisit with a new 
perspective the experimental setup presented in Ref.~\onlinecite{BernierKoehl2009}. We further 
highlight the differences between our cooling method and another one recently proposed 
in Ref.~\onlinecite{HoZhou2010}.

\subsubsection{Idealized realization}
\begin{figure} [bt]
    \centerline{
    \includegraphics[width=8cm,clip=true]{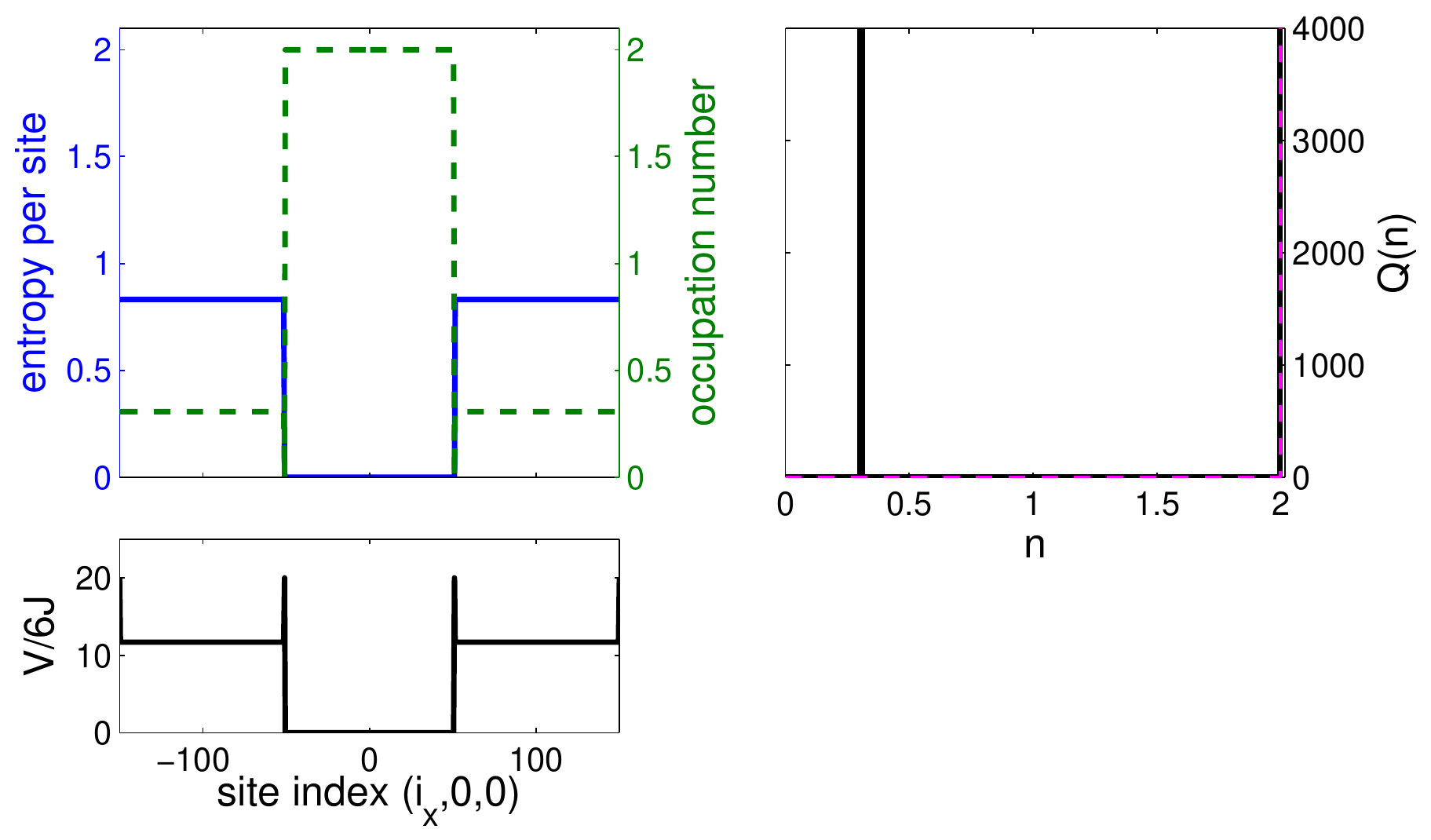}
    }
    \caption{(Color online) Occupation number (dashed line, left upper panel), entropy per particle (solid line, left upper panel),
             and potential profile (solid line, lower panel) for the idealized trapping potential. $Q(n)$ (solid line (full system),
             dashed line (core region), right upper panel) as a function of the density. The core region is taken to be cylindrical
             with $r_0 = 51a$, the storage region is a rectangular box of size $300 \times 300 \times 100 a^3$. Each region has a
             homogeneous density $n_\text{core} = 2$, $n_\text{storage} = 0.31$. The average entropy per particle for the total system
             is $\frac{S_T}{N_T} = 1.65$ and $\frac{S_C}{N_C} = 0$ for the core region. The ratio of particles
             in the core region versus the total particle number is $\frac{N_C}{N_T} = 0.39$.}
    \label{fig:idealized_trap}
\end{figure}

As explained above, we first present our cooling method using an idealized setup uncluttered with experimental
details. In the following, we show how, given an idealized two-fluid model, one can start with a gas, loaded in an
optical lattice and confined to an harmonic trap, having an entropy per particle, e.g.,around $\frac{S}{N}|_\text{initial} = 1.7$ and
obtain a final system characterized by a near zero entropy per particle while keeping about $40\%$ of the atoms.

As we have seen in Fig.~\ref{fig:S_over_n_n_hom}, at high temperatures, the entropy per particle $\frac{s(n)}{n}$ is largest for
low densities. Therefore to segregate the entropy in our system, we would like to
create two distinct regions (Fig.~\ref{fig:idealized_trap}): (i) a ``core region''
with a deep trapping potential in which the density $n_C$ is close to two particles per site, i.e.~$n_C\approx 2$ and
(ii) a ``storage region'' with a very flat trapping potential in which the density $n_S$ is very low, i.e.~$n_S\ll 1$.
In the language of the $Q(n)$ distribution introduced in the previous sections, this idealized setup results in
two sharp peaks: one at a very low density due to the storage region and
another one very close to $n=2$ due to the core.

To obtain these two regions of unequal densities, we start from a gas confined to the usual harmonic trapping
potential with total entropy $S_T$, and adiabatically deform the trap to reach the trapping profile presented in
Fig.~\ref{fig:idealized_trap}. While the total entropy remains constant under this deformation, the entropy is now
inhomogeneously distributed. The entropy per particle in the core region will be very low, ideally zero, whereas in the
storage region the entropy per particle will be quite large. In contrast, the temperature $T_0$ remains equal throughout
the system and is set by the constraint requiring entropy conservation, i.e.~
\begin{equation}
\label{eq:entropy_shaping}
S = \int_0^2 dn \left\{ Q_C(n) \frac{s(n,T_0)}{n} + Q_S(n) \frac{s(n,T_0)}{n} \right\} \equiv S_T
\end{equation}
where $Q_C(n) = \sum_\text{i $\in$ C} n_i~\delta(n_i-n)$ and $Q_S(n) = \sum_\text{i $\in$ S} n_i~\delta(n_i-n)$.
For the idealized two-fluid model presented here, Eq.~\ref{eq:entropy_shaping} simplifies to
\begin{equation}
\label{eq:entropy_shaping2}
S = N_C \frac{s(n_C,T_0)}{n_C} + N_S \frac{s(n_S,T_0)}{n_S} \equiv S_T
\end{equation}
where $N_C$ and $N_S$ are the number of particles in the core and storage regions respectively.

In this idealized situation an infinitely narrow barrier is used to separate the two regions adiabatically.
After the separation, the high entropy region can be removed and the core region can be used to perform the experiment.
As the density is uniform through out the core region, the average entropy per particle at the
time of separation is given by $\frac{S_\text{final}}{N_\text{final}} = \frac{s(n_C,T_0)}{n_C}$. Therefore, the final
average entropy per particle in the core region, $\frac{S_\text{final}}{N_\text{final}}$,
can be much lower than the initial average entropy per particle. It is important to note that after this separation
further adiabatic changes of the system parameters can lead to temperature changes, but the average entropy per
particle will remain constant.

The minimal entropy which can be reached depends on different quantities as seen from
Eq.~\ref{eq:entropy_shaping2}. As expected the initial entropy $S_T$ is an important factor.
Furthermore, the cooling procedure is more efficient if the entropy per particle in the storage region is much
larger than the entropy per particle in the core region, i.e.~$\frac{s(n_C,T_0)}{n_C} \ll \frac{s(n_S,T_0)}{n_S}$.
For fixed temperatures, a typical behavior of the entropy per particle with the density is shown
in Fig.~\ref{fig:S_over_n_n_hom}. As one can see the decrease in entropy per particle with increasing
density is more pronounced at larger temperatures. Hence the procedure will become less efficient at lower initial
temperature. 

We also find that the differences in entropy per particle are largest between very low and very high densities,
such that the procedure would be most efficient if such densities were used for the storage and core regions respectively.
However, it is important to note that it is not essential that the state stabilized in the core region is gapped
as the band insulator is. What really matters is that a sizable difference is achieved between the entropies per
particle characterizing the densities of the core and storage regions. By tuning the trap shape the
number of atoms in the core and storage regions can be adjusted. Ideally one would like to create a very
large storage region with a lot of atoms at very low density, since there the entropy per atom would be maximal.
However, this situation can only be achieved within a certain range due to experimental limitations. In particular
the trap can only be engineered within a certain spatial extension and the final number of particle ($N_\text{final} = N_C$)
should be reasonably large in order to perform the actual experiment afterward.

Before turning our attention to the experimentally realizable setup, one important comment is in order.
We want to emphasize that the adiabatic isolation of the core region from the storage region is favorable or even
necessary to achieve cooling. Without proper isolation, the entropy of the storage region can flow back into the
experimentally relevant region. At best this backflow of entropy may simply render the cooling scheme
less efficient, but in the worst scenario it may heat up the region of interest. At low temperature, heating may occur
because the experimentally relevant phase may accommodate more easily the entropy than its parent high density state
(in our case the band insulator). Therefore, while changing back the trap shape to generate the experimentally
relevant phase, the entropy may flow back into the region of interest if the core and storage regions are not
properly isolated from each other.

\subsubsection{Realistic realization}
\begin{figure} [bt]
    \centerline{
    \includegraphics[width=8cm,clip=true]{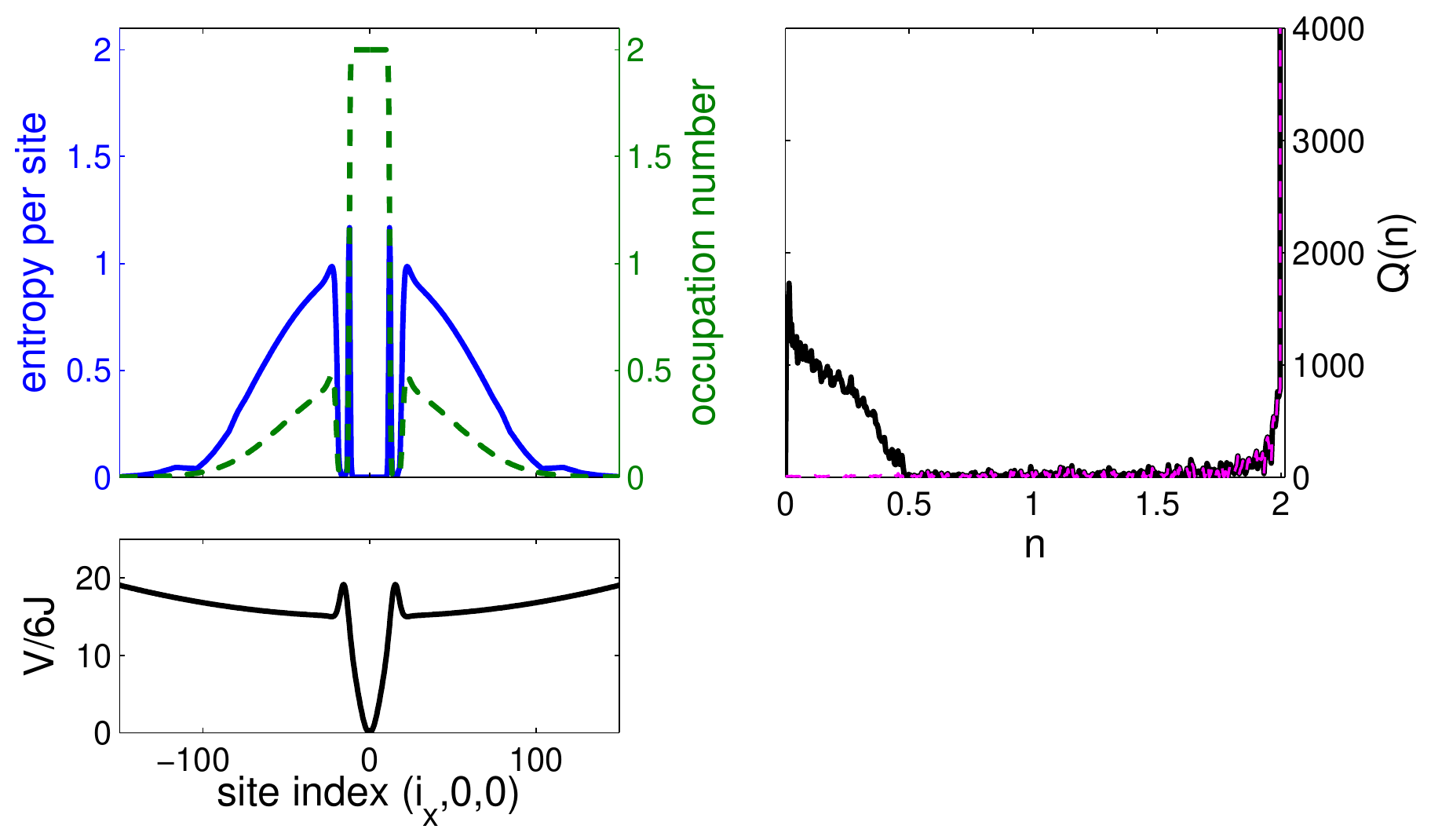}
    }
    \caption{(Color online) Occupation number (dashed line, left upper panel), entropy per particle (solid line, left upper panel),
             and potential profile (solid line, lower panel) in the presence of the dimple and barriers, as a function of the
             transverse coordinate. The potential is offset such that V = 0 is at the bottom of the dimple. $Q(n)$ (solid line (full system),
             dashed line (core region), right upper panel) as a function of the density. We chose the following experimentally
             realistic parameters: $\frac{U}{6J} = 0.5$, $\frac{V_h}{6J} = 1.8 × 10−4$, $\gamma^2 = 50$, $\frac{V_b}{6J} = 6$, $r_b = 15a$,
             $\omega_b = 5a$, $\frac{V_d}{6J} = 15$, $\omega_d = 15a$, and $12 \cdot 10^4$ atoms. The average entropy
             per particle in the total system is $\frac{S_T}{N_T} = 1.95$ and in the core region $\frac{S_C}{N_C} = 0.198$.
             The ratio of particles in the core
             region versus the total particle number is $\frac{N_C}{N_T} = 0.404$. Obtained with DMFT. See also Fig. 2 of Ref.~\onlinecite{BernierKoehl2009}.}
    \label{fig:realistic_trap}
\end{figure}

Let us now turn our attention to the experimentally realizable trapping potential.
To be truly relevant, this trapping profile should be achievable using present technology and
should not require fine-tuning of the system parameters. Such an experimentally realistic trapping
potential has been described in Ref.~\onlinecite{BernierKoehl2009}. Here we only briefly summarize the setup.
In order to achieve the required entropy modulation, the potential
has to allow for a tight trapping in the core region, for a wide shallow ring in the storage region and
for high potential barriers isolating these two regions from each other. To produce this profile,
we envision to use three elements, (i) a shallow harmonic trap (either magnetic or optical), (ii) a dimple
which confines atoms to a small region around the trap symmetry axis and helps to create the band insulator,
and (iii) a cylindrically-symmetric potential barrier to isolate high and low entropy regions. The
dimple (ii) and potential barrier (iii) are produced by red- and blue-detuned laser beams respectively,
creating attractive or repulsive dipole potentials. The dimple has a Gaussian profile, while the barrier
needs to be a narrow annulus. Experimentally this can be realized either by setting a tightly focused laser beam
in rapid rotation, or by engineering the beam profile using phase plates or other diffractive
optics\cite{Friedman2002}. Consequently, in addition to the lattice potential, the trapping profile
is given by
\begin{eqnarray}
&&V({\bf r}) = V_\text{harmonic} + V_\text{dimple} + V_\text{barrier} \nonumber
\\
 \textrm{with}
&& V_\text{harmonic}({\bf r}) = V_\text{h}~(x^2+y^2+\gamma^2 z^2)/a^2, \nonumber \\
&& V_\text{dimple}({\bf r}) = -V_\text{d}~\exp{(-2(x^2+y^2)/w_\text{d}^2)}. \nonumber \\
&& V_\text{barrier}({\bf r}) = V_\text{b}~\exp{(-2(\sqrt{x^2+y^2}-r_\text{b})^2/w_\text{b}^2)}, \nonumber
\end{eqnarray}
where $V_{\{\text{h,d,b}\}}$ are the potential amplitudes, $\gamma$ is a measure of the anisotropy
of the harmonic trap, $w_{\{\text{d,b}\}}$ are the waists of the Gaussian laser beams forming the dimple and
barrier, $r_b$ is the radius of the cylindrical barrier, and $a$ the lattice spacing.
An example of the resulting trapping profile is shown in Fig.~\ref{fig:realistic_trap}. The gain
in entropy per particle for this particular case is approximately one order of magnitude while keeping
about 50$\%$ of the atoms. In Fig.~\ref{fig:realistic_trap}, in addition to the density distribution and
entropy per particle for the corresponding trapping profile, one can see on the right panel from the
$Q(n)$ distribution that the two sharp peaks of the idealized setup have been replaced with broader
features. In particular, the lower peak is not well defined and removing the storage region
results in an almost complete extinction of its weight.

However, experimentally removing the storage region is not straightforward due to the presence of the optical lattice
potential. In Ref.~\onlinecite{BernierKoehl2009}, several possible methods have been proposed. Fortunately, since
the publication of this article, different techniques aiming at locally addressing the atoms have been developed
and could be employed to remove the storage region. One of these methods makes use of an
electron microscope and could be used to blast away the storage region in a very controlled manner \cite{WuertzOtt2009}. Another
possibility would be to use locally trapped ions to remove the atoms from the storage region \cite{ZipkesKoehl2010}.

All previous considerations assumed the system evolution to be perfectly adiabatic. In a real 
experiment this will not be the case as the deformation of the trap needs to be performed 
within a finite time. In Ref.~\onlinecite{BernierKoehl2009}, we showed that, for a one-dimensional system, 
reshaping the trapping potential can be performed within an experimentally realistic timescale 
while inducing heat on a scale more than ten times smaller than the antiferromagnetic exchange coupling.

Consequently, using the scheme presented above, experimentalists could cool down quantum gases to
one order of magnitude lower than presently achievable while keeping about half the atoms in the system.
Cooling into highly sought-after quantum phases could thus be achieved.

\section{Conclusion}

In this article, we performed a detailed study of the thermodynamics of the
three-dimensional fermionic Hubbard model, for a rather wide range of couplings. We
mainly focused on the temperature regime $k_B T/6J \gtrsim 0.1$, of current interest
for experiments on cold fermionic atoms in optical lattices.

Our theoretical study is mainly based on single-site dynamical mean-field theory (DMFT), a
well established theoretical method based on a controlled approximation
in which non-local correlations are neglected but local quantum fluctuations
are treated accurately. In addition, we used high-temperature series
expansion. A comparative study between these two methods allowed for a
precise assessment of their respective range of validity. DMFT is found to be
accurate down to fairly low temperatures when not too close to half-filling (one
particle per site). Because of their convergence properties, the series expansion
are most useful at low density or exactly at half-filling.
At or close to half-filling, single-site DMFT can be trusted only down
to $T\sim J$. Below this temperature, short-range magnetic correlations set in, which
require the use of cluster extensions of DMFT.
Hence, our study validates the use of single-site DMFT for understanding experimental
results on cold fermionic atoms in a 3D lattice in the currently accessible temperature
range, while future experiments at lower temperature will require extensions of the
method.

We considered the implications of the thermodynamic properties of the homogeneous system
for fermionic atoms in an optical lattice confined in a parabolic potential, within the
local density approximation. A state
diagram was established, with different regimes for the density profile in the trap.
Special emphasis was devoted to the distribution function of site occupancies in the
trap, a quantity which can be experimentally measured by imaging techniques with
single lattice-site resolution. Such techniques have recently become available
for two-dimensional systems.

This distribution function proved to be of particular use when discussing how
a given observable (e.g., temperature) changes in the trap as the system evolves in an adiabatic manner when
a parameter is varied.
Indeed, it allows for a clear separation of two contributions, one
corresponding to the redistribution of atoms in the trap during the evolution,
and another from the intrinsic change of the observable.
We applied these considerations to the temperature change in the trap under an adiabatic
variation of the coupling, and identified the regimes where a significant
cooling can be expected.
%
However, we would like to point out that for these results to be completely applicable 
to trapped cold atoms confined to optical lattices, one would need to take into account several
processes that could prevent a fully adiabatic evolution. 
Considering the effects of processes such as lattice heating will be the subject of future works. 

Finally, we elaborated on a previously proposed cooling mechanism, based on the
separation of regions of small and large entropy.
This procedure is promising for cooling further cold fermionic systems
by approximately one order of magnitude, a necessary step and major future
challenge for
accessing and studying experimentally strongly correlated phases in those systems.

{\it Note added:} When the present work was near completion we became aware that 
another study, conducted by Fuchs {\it et al.}, on the thermodynamics of the 3D
Hubbard model was also in its final stage. The focus of Ref.~\onlinecite{FuchsTroyer2010} 
is on the characterization of the low temperature region near the 
antiferromagnetic transition. In the temperature range where our two works 
overlap, the results are in agreement.

\acknowledgements
We are grateful to M.~Ferrero and O.~Parcollet for discussions and
for sharing with us their CT-QMC code. We thank S.~Wessel for providing us with the entropy 
data of the Heisenberg model from Ref.~\onlinecite{Wessel2010}. We
acknowledge discussions and previous collaborations on related subjects with
I.~Bloch, T.~Esslinger, F.~Gerbier, E.~Gull, R.~J\"{o}rdens,
M.~K\"{o}hl, J.~Oitmaa, L.~Pollet, A.~Rosch, C.~Salomon, and M.~Troyer.
We acknowledge funding from the Agence Nationale de la Recherche,
the DARPA-OLE program, the `Triangle de la Physique' and the Fonds Qu\'eb\'ecois
de Recherche sur la Nature et les Technologies.

\appendix
\section{A simple analytical approximation}
\label{sec:simpleapprox}

In this appendix, we present a very simple approximation that allows us 
to obtain the state
diagram (Fig.~\ref{fig:state_diagram} and Ref.~\onlinecite{DeLeoParcollet2008})
of trapped fermionic atoms in an optical lattice with very little
computational effort.
It also allows us to
obtain approximate analytical expressions for the crossover lines
in this state diagram, hence providing qualitative understanding into the
numerical results presented above in this article.

\begin{figure}
    \includegraphics[width=7cm,clip=true]{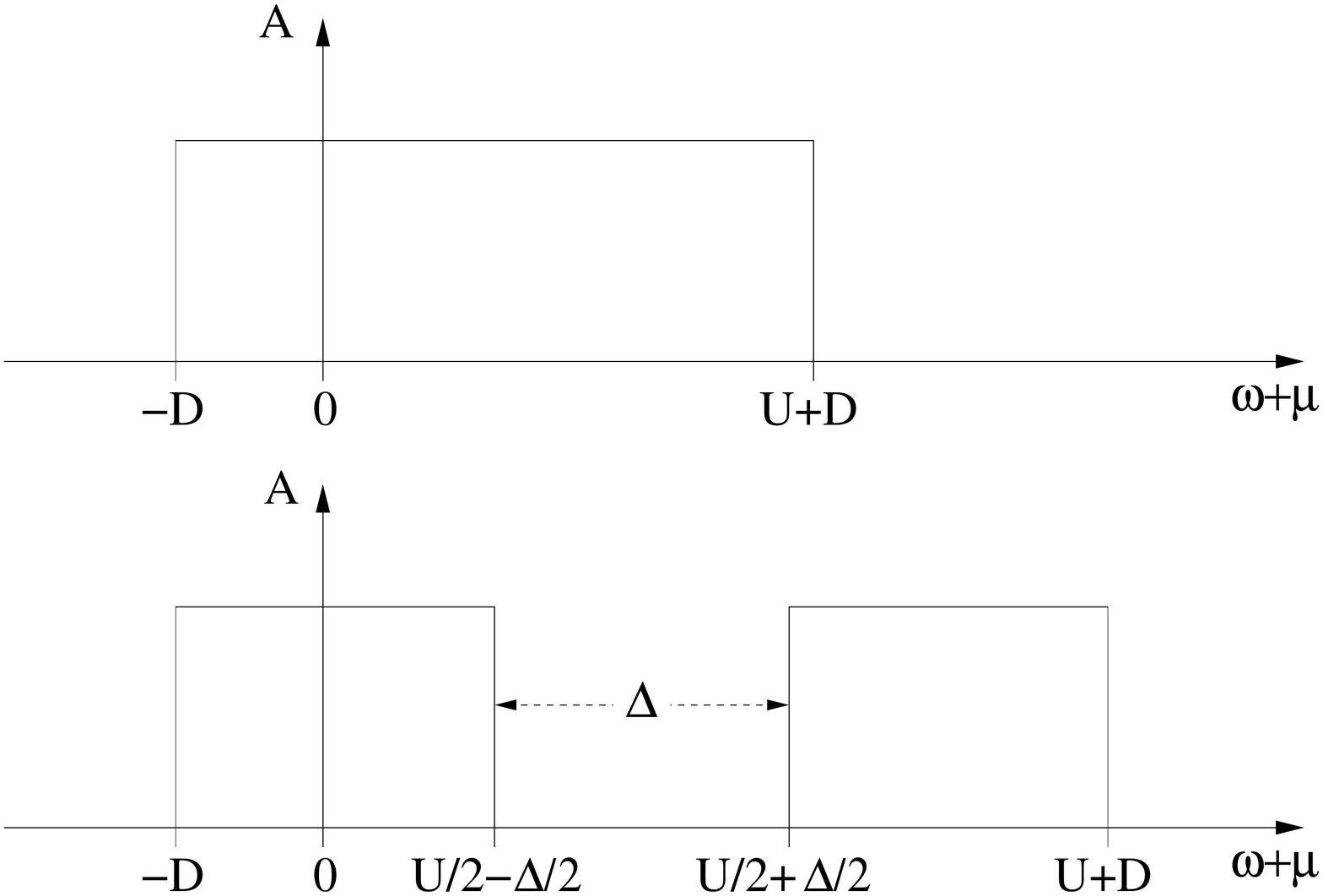}
    \caption{Modelization of the spectral function in the metallic (above) and
    insulating (below) phases. In the metallic phase $\Delta=0$, while in the
    insulating phase $\Delta > 0$.
    }
    \label{fig:analytic_dos}
\end{figure}

This approximation is based on the following approximate form of the one-particle
spectral function of the homogeneous system:
\begin{equation}
    A(\omega) = \left\{
    \begin{array}{cl}
    0 & \qquad \text{for}\,\, \omega+\mu < -D \\
    \frac{1}{U-\Delta+2D} & \qquad \text{for}\, -D < \omega+\mu < \frac{U}{2}
        - \frac{\Delta}{2} \\
    0 & \qquad \text{for}\, \frac{U}{2} - \frac{\Delta}{2} < \omega+\mu <
        \frac{U}{2} + \frac{\Delta}{2} \\
    \frac{1}{U-\Delta+2D} & \qquad \text{for}\, \frac{U}{2} +
        \frac{\Delta}{2} < \omega+\mu < U + D \\
    0 & \qquad \text{for}\, \omega+\mu > U+D.
    \end{array} \right.
\end{equation}
In this expression, $D\equiv 6J$ is the half-bandwidth and
$\Delta$ is a parameter that plays the role of the Mott gap (see
below). The rationale behind this expression is the following. When $\Delta=0$,
(weakly interacting regime) it describes a density of states broadened by interaction
effects. When $\Delta\neq 0$, it describes two Hubbard bands (Fig.~\ref{fig:analytic_dos})
separated by a Mott gap. The contribution of quasiparticle states
appearing in between the two Hubbard bands are neglected, because the
temperatures considered in this article are typically larger than the quasiparticle
coherence scale.

Compared to the atomic limit, this approximation has a better behavior in the low temperature
limit while still retaining a simplicity that allows for a completely analytic solution.
The disadvantage of the atomic limit is that it models the zero temperature spectral 
function as an unrealistic pair of delta functions located at $0$ and $U$. The atomic limit is indeed the first
term of the expansion in $\beta J$ and as such is valid at high temperature.
Our approximation instead takes into account the fact that the Hubbard bands
are broadened by the kinetic term. The result is in better agreement with the
DMFT data at temperatures lower than the limit of validity of the atomic limit.
Our approximation eventually breaks down due to the fact that at low
temperature the details of the shape of the Hubbard bands become more relevant
and the box-like model of the bands shows its limits. In this sense,
our approximation should only be regarded as acceptable at high
temperature.

This form of the spectral function leads to the following expression of the
dependence of the particle number on chemical potential:
\begin{eqnarray}
n(\mu,T)&=&2 \int_{-\infty}^\infty d\omega A(\omega) f(\omega)\\ \nonumber
&=& 1 + \frac{2k_BT}{U-\Delta+2D} \ln \left\{ \frac
    {\cosh \left( \frac{\mu + D}{2k_BT} \right)}
    {\cosh \left( \frac{\mu - U - D}{2k_BT} \right)} \cdot \right. \nonumber \\
    && \left. \cdot \frac
    {\cosh \left( \frac{\mu - \frac{U}{2} - \frac{\Delta}{2}}{2k_BT} \right)}
    {\cosh \left( \frac{\mu - \frac{U}{2} + \frac{\Delta}{2}}{2k_BT} \right)}
    \right\}. \label{eq:n_analytic_model}
\end{eqnarray}
In the zero-temperature limit, this expression reduces to:
\begin{eqnarray}
n(\mu)&=&0 \,\, (\mu<-D)\nonumber \\
&=&\frac{2}{U-\Delta+2 D} (\mu+D) \,\, \mathrm{for}\,\,-D<\mu<\frac{U-\Delta}{2}\nonumber \\
&=& 1 \,\, \mathrm{for}\,\,\frac{U-\Delta}{2}<\mu<\frac{U+\Delta}{2}\nonumber\\
&=& 1+\frac{2}{2D+U-\Delta}(\mu-\frac{U+D}{2})\,\,\\
&\mathrm{for}&\,\,\frac{U+\Delta}{2}<\mu<D+U\nonumber\\
&=2&\,\,\mathrm{for}\,\, \mu>D+U
\end{eqnarray}
i.e. a function with steps at $n=0$, $n=1$ and $n=2$ (zero compressibility),
while $n(\mu)$ in between the steps (liquid regions) is approximated by a
linear dependence on $\mu$ (constant compressibility approximation).
%

\begin{figure}
    \includegraphics[width=7cm,clip=true]{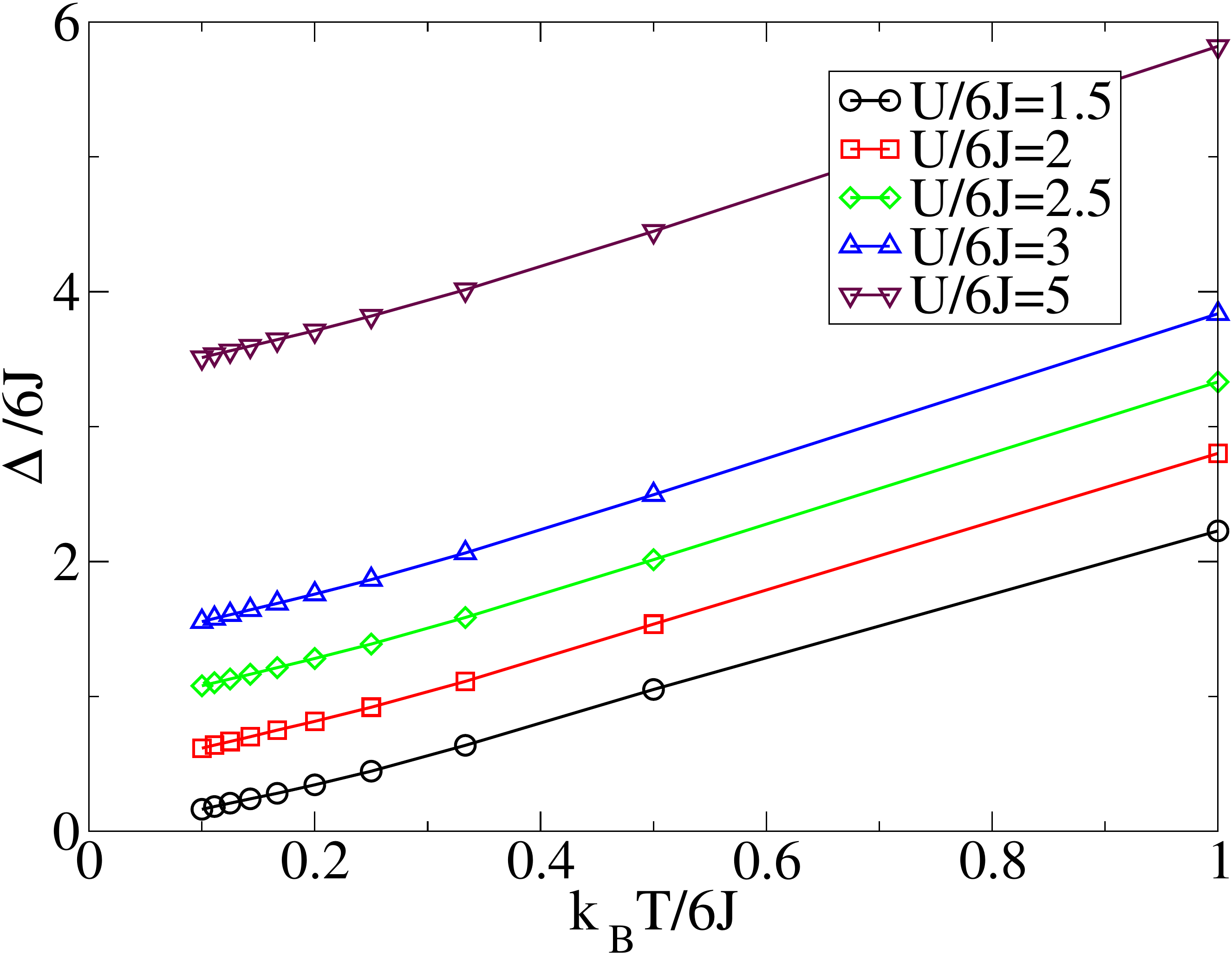}
    \caption{The gap parameter $\Delta(\mu,T)$ as a function of $T$ for different interaction strength $U$.
    }
    \label{fig:gap_vs_T}
\end{figure}

We regard the parameter $\Delta$ as a fitting parameter, function of coupling and
temperature, and perform a least square minimization of Eq.~\ref{eq:n_analytic_model}
to the DMFT data to determine its value.
The result of this procedure is displayed in
Fig.~\ref{fig:gap_vs_T}. The fit looses in quality as we reduce the
temperature and the value of $U$. Restricting to the high temperature and large
$U$ region, the dependence of $\Delta$ on $U$ and $T$ can be separated to a
good approximation and $\Delta(U,T)$ can be fit remarkably well by the law:
\begin{equation}
    \Delta(U,T) \sim \Delta_0 + \alpha_U U + \alpha_T k_BT = -1.7 + 0.96 U + 2.66 k_BT
\end{equation}
where the parameter with the largest deviation is $\alpha_T$ which shows a
slight increase as a function of $U$.

\begin{figure}[ht!]
    \includegraphics[width=7cm,clip=true]{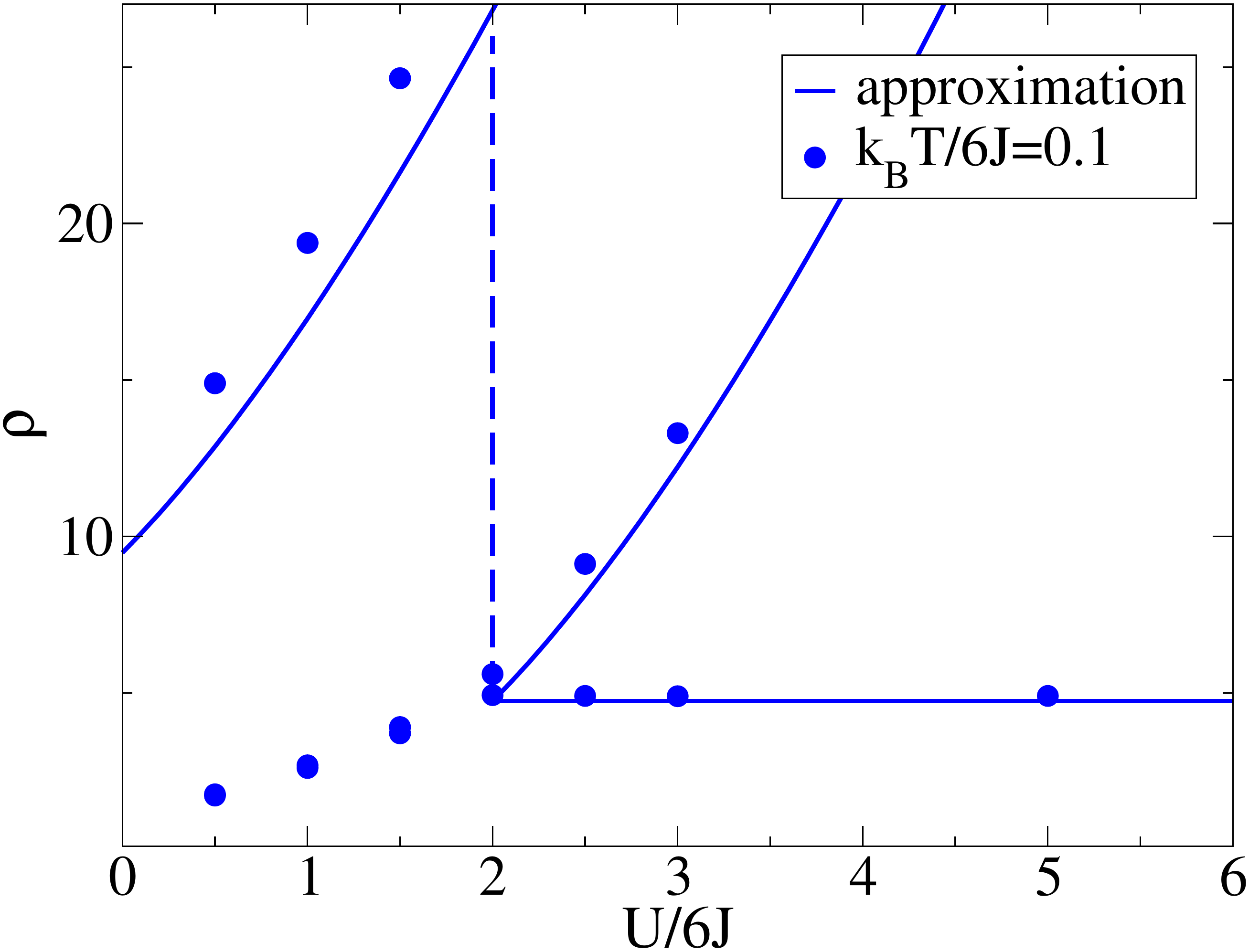}
    \caption{Comparison of the state diagram obtained by DMFT (circles)
    and the simple analytic approximation (solid lines). [cf. Ref.~\onlinecite{DeLeoParcollet2008}]
    }
    \label{fig:state_diagram_app}
\end{figure}

Using the approximate expression (\ref{eq:n_analytic_model}) of $n(\mu,T)$
for the homogeneous system, it is straightforward to obtain the state diagram of the
trapped system by using the relation between the scaled density $\rho$ and the
chemical potential $\mu_0$ at the center of the trap, which reads,
in the LDA approximation:
\begin{equation}
\rho\equiv N \left(\frac{V_t}{D}\right)^{3/2}=
2\pi \int_{-\infty}^{\mubar_0} d\mubar \sqrt{\mubar_0-\mubar}\, n(\mubar,T).
\label{eq:lda_bis}
\end{equation}
The calculations can be performed analytically in the $T=0$ limit (we
warn the reader that $T=0$ is considered here only as a formal limit
in order to make
an analytical calculation possible, since the approximations made in this
paper are no longer valid physically in this limit).
Let us for example focus on the crossover lines which delimit
the `Mc' region in Fig.~\ref{fig:state_diagram}) within which
the central occupancy is $n=1$.
In view of our approximate form of $n(\mu)$, the lower boundary of this
regime will correspond to $\mu_0=(U-\Delta)/2$ and the upper boundary to
$\mu_0=(U+\Delta)/2$. Inserting these values into (\ref{eq:lda_bis}) and performing the
integrations at $T=0$ yields the following expressions for the lower and
upper boundaries of the Mc region:
\begin{equation}
\rho_{Mc}^{<} = \frac{2\sqrt{2}\pi}{15}\,(\frac{U-\Delta}{D}+2)^{3/2}
\label{eq:n=1inf_U>Uc}
\end{equation}
\begin{equation}
\rho_{Mc}^{>} = \frac{2\sqrt{2}\pi}{15}\,\frac{(\frac{U+\Delta}{D}+2)^{5/2}-(2\Delta/D)^{5/2}}
{(U-\Delta)/D+2}.
\label{eq:n=1sup_U>Uc}
\end{equation}
Using the above determination of $\Delta$, this provides an explicit form of the
boundaries.
Analytical expressions can be similarly obtained for all crossover lines in the state
diagram. In Fig.~\ref{fig:state_diagram_app} we compare these approximate analytical expressions
to the actual lines obtained for a DMFT calculation at a low enough temperature $T/6J=0.1$
and find very satisfactory agreement.

Finally, we note that thermodynamic quantities such as the double occupancy and
the entropy can in principle be reconstructed from a given approximate expression for
$n(\mu,T,U)$ thanks to the thermodynamic (Maxwell) relations relating their derivatives:
\begin{eqnarray}
 \frac{\partial s}{\partial \mu} &=& \frac{\partial n}{\partial T}
    \label{eq:thermodyn_2}\\
    \textrm{and }\quad
    \frac{\partial d}{\partial \mu} &=&
    -\,\frac{\partial n}{\partial U}.
    \label{eq:thermodyn_3}
\end{eqnarray}


\end{document}